\documentclass{aa}
\pdfoutput=1
\usepackage{graphicx}
\usepackage{epsfig}
\usepackage{txfonts}
\usepackage{wasysym}
\usepackage{marvosym}

\def\mearth{M_\oplus}
\def\msun{M_\odot}
\def\rhill{R_{\rm H}}
\def\mcore{M_{\rm core}}
\def\mheavy{M_{\rm Z}}
\def\astart{a_{\rm start}}
\def\fpg{f_{\rm D/G}} 
\def\fri{f_{\rm R/I}}
\def\f1{f_{\rm I}}
\def\mj{M_{\textrm{\jupiter }}}
\def\mstar{M_*}
\def\mdisk{M_{\rm disk}}
\def\tdisk{t_{\rm disk}}
\def\astart{a_{\rm start}}
\def\tstart{t_{\rm start}}
\def\membstart{M_{\rm emb,0}}
\def\memb{M_{\rm emb}}
\def\anorm{a_0}
\def\rcore{R_{\rm core}}
\def\rcapt{R_{\rm capt}}
\def\mdotcore{\dot{M}_{\rm core}}
\def\mdotst{\dot{M}_{\rm *}}
\def\mwind{\dot{M}_{\rm w}}
\def\menv{M_{\rm env}}
\def\miso{M_{\rm iso}}
\def\mpla{m_{\rm pla}}
\def\arock{a_{\rm rock}}
\def\aice{a_{\rm ice}}
\def\beq{\begin{equation}}
\def\eeq{\end{equation}}
\def\mplanet{M_{\rm planet}}
\def\aplanet{a_{\rm planet}}
\def\sigmad{\Sigma_{\rm D}}
\def\atouch{a_{\rm touch}}
\def\sigmanorm{\Sigma_{0}}

\def\nsynt{N_{\rm synt}}

\def\msini{M\sin i}
\def\amax{a_{\rm max}}
\def\amin{a_{\rm min}}
\def\aphys{a_{\rm phys}}

\def\mmax{M_{\rm max}}

\def\aj{AJ}                   
\def\araa{ARA\&A}             
\def\apj{ApJ}                 
\def\apjl{ApJ}                
\def\apjs{ApJS}               
\def\ao{Appl.Optics}          
\def\aap{A\&A}                
\def\mnras{MNRAS}             
\def\pasp{PASP}               

\begin{document}


\title{Extrasolar planet population synthesis I:\\ Method, formation tracks and mass-distance distribution}

\author{Christoph Mordasini\inst{1,2}  \and Yann Alibert\inst{1,2}  \and Willy Benz \inst{1}}

\institute{Physikalisches Institut, University of Bern, Sidlerstrasse 5, CH-3012 Bern, Switzerland \and
Current address: Max-Planck-Institut f\"ur Astronomie, K\"onigstuhl 17, D-69117 Heidelberg, Germany \and
Institut UTINAM, CNRS-UMR 6213, Observatoire de Besan\c{c}on, BP 1615, 25010 Besan\c{c}on Cedex, France } 

\offprints{Christoph MORDASINI, \email{mordasini@mpia.de}}

\date{Received 1 June 2008  / Accepted 15 April 2009}

\abstract
{With the high number of extrasolar planets discovered by now, it becomes possible to use the properties of this planetary population to constrain theoretical formation models in a statistical sense. This paper is the first in a series in which we carry out a large number of planet population synthesis calculations within the framework of the core accretion scenario. We begin the series with a paper mainly dedicated to the presentation of our approach, but also the discussion of a representative synthetic planetary population of solar like stars. In the second paper we statistically compare the subset of detectable planets to the actual extrasolar planets. In subsequent papers, we shall extend the range of stellar masses and the properties of protoplanetary disks.}
{The last decade has seen a large observational progress in characterizing both protoplanetary disks, and extrasolar planets. Concurrently, progress was made in developing complex theoretical formation models. The combination of these three developments allows a new kind of studies: The synthesis of a population of planets from a model,  which is compared with the actual population. Our aim is to get a general overview of the population, to check if  we quantitatively reproduce the most important observed properties and correlations, and to finally make predictions about the planets that are not yet observable.}
{Based as tightly as possible on observational data,  we have derived probability distributions for the most important initial conditions for the planetary formation process. We then draw sets of initial conditions from these distributions and obtain the corresponding synthetic planets with our formation model. By repeating this step many times, we synthesize the populations.}
{Although the main purpose of this paper is the description of our methods, we present some key results:  We find that the variation of the initial conditions in the limits occurring in nature leads to the formation of planets of large diversity. This formation process is best visualized in planetary formation tracks in the mass-semimajor axis diagram, where different phases of concurrent growth and migration can be identified.  These phases lead to the emergence of sub-populations of planets distinguishable in a mass-semimajor axis diagram. The most important ones are the ``failed cores'', a vast group of core-dominated low mass planets, the ``horizontal branch'', a sub-population of  Neptune mass planets extending out to 6 AU, and the ``main clump'', a concentration of giant gaseous giants planets at around 0.3-2 AU.}
{}


\titlerunning{Extrasolar planet population synthesis I}
\authorrunning{C. Mordasini et al.}

\maketitle

\section{Introduction}\label{sect:introduction}
As of spring 2009, more than 300 extrasolar planets have been discovered (J. Schneider's Extrasolar Planet Encyclopedia at http://exoplanet.eu). The richness and diversity of the characteristics of these exoplanets like their mass or semimajor axis is impressive, and was not necessarily expected from the single example - our own solar system - that was available to study  before the discovery of 51 Peg b (\object{HD\,217014b}) by Mayor \& Queloz (\cite{mayorqueloz1995}). 

Since then, the observational field of extrasolar planet search has seen a rapid evolution leading to numerous additional discoveries of planets orbiting other stars. These discoveries have also triggered numerous theoretical  studies about the formation and evolution of these planets. Key physical processes in planet formation and evolution could be identified whose importance was not fully realized in previous works based on the solar system alone.

Some of these discovered planets, and multiple planetary systems, are sufficiently interesting by themselves to warrant individual theoretical studies. Examples are the extrasolar planetary system with three Neptune-mass planets around \object{HD\,69830} (Lovis et al. \cite{lovisetal2006}; Alibert et al. \cite{alibertnewsyst2006}), or the  transiting Neptune mass planet \object{GJ\,436b} (Butler et al. \cite{butleretal2004}; Gillon et al. \cite{gillonetal2007}; Figueira et al. \cite{figueiraetal2008}). Of course, the giant planets in our own solar system provide a much larger and detailed set of constraints than any known extrasolar planet. Therefore, each formation model applied to discuss extrasolar planet formation should also be put to the test to reproduce the characteristics of our own giant planets (Pollack et al. \cite{pollacketal1996}, Alibert et al. \cite{alibertetal2005b}; Hubickyj et al. \cite{hubickyjetal2005}; Benvenuto \& Brunini \cite{benvenutobrunini2005}).   

The modeling of the formation of such single systems while a necessary condition to validate formation models is not satisfactory by itself. Indeed, the number of model parameters is generally large while the  number of constraints deriving from a single system is small, and not strong enough to completely constrain any formation model.

Thanks to the rapid growth of the number of known extrasolar planets, the situation has however dramatically changed: Instead of having only a single object or a single system to study, we now begin to be able to describe an entire \textit{population} of extrasolar planets orbiting FGK stars in the solar neighborhood. While this population is still smaller than one would ideally like, it nevertheless already allows to extract statistically a wealth of information (e.g. Udry \& Santos \cite{udrysantos2007}; Cumming et al. \cite{cummingetal2008}) to constrain formation models that exceeds by far what one extrasolar planet can do.  This is especially true since most of the extrasolar planets have been discovered by radial velocity measurements so that only a few orbital elements and a minimum mass are known for one individual object. For the growing number of transiting planets more physical properties can be derived and compared with internal structure models (Baraffe et al. \cite{baraffeetal2008}; Figueira et al. \cite{figueiraetal2008}).  Unfortunately, transiting planets known so far are all in close proximity to their host star.  Hence it is sometimes unclear to what extend their characteristics are still related to their formation or rather to subsequent evolution (e.g. evaporation). 

Parallel to the discovery of more and more end-products of the planetary formation process \textit{i.e.} planets, large observational progress (e.g. Meyer et al. \cite{meyeretal2006}) has also been made in characterizing the initial conditions for this process, \textit{i.e.} the protoplanetary disks. Thanks to these observations, we begin to be able to determine the probability of occurrence of any particular initial condition for planetary formation, like disk metallicity, mass or lifetime. 

With these two sets of observational data at hand, a new interesting class of theoretical planet formation studies has become possible, where a theoretical model serves as the link between these two groups of observations: The synthesis of populations of planets by Monte Carlo methods. In this approach the observed distributions of disk properties are used as varying initial conditions for the model. The final characteristics of the synthetic planets that form in the model can then be compared statistically to those of the actual observed populations.  This addresses the question if the observed diversity of extrasolar planets is simply the consequence of the diversity of disk properties. 

As we shall show, such studies have proven to be very fruitful, as they not only allow to reproduce observations but also show the links and correlations between the different initial conditions and the characteristics of the resulting planets. Thereby they provide great insights into the formation mechanism. Last but not least, such an approach by predicting the actually existing planet population as opposed to the actually detected one, allows to optimize future searches and instruments when coupled to a synthetic detection bias for a particular detection method.
 
Compared to similar studies e.g. the pioneering work of Ida \& Lin (\cite{idalin2004a}, \cite{idalin2004b}, \cite{idalin2005} and \cite{idalin2008}), or the studies of Kornet \& Wolf (\cite{kornetwolf2006}), Robinson et al. (\cite{robinsonetal2006}) and Thommes et al. (\cite{thommesetal2008}), we have attached particular importance to three distinct areas: First, we use the detailed extended core accretion formation model of Alibert et al. (\cite{alibertetal2005a}) that has been successfully applied to quantitatively explain the many observed constraints of the giant planets of our own solar system (Alibert et al. \cite{alibertetal2005b}). Second, we stick as tightly as possible to probability distributions derived from observations,  and third (cf. the companion paper Mordasini et al. (\cite{paperII}), hereafter paper II), we use quantitative statistical methods to compare model outcomes and observations and require that as many different observational constraints as possible be satisfied at the \textit{same} time. In this way, we can check which observed properties can be reproduce by our formation model. But also discrepancies between the synthetic and the actual population provide new insights allowing to improve the models. 

In this first aper, we present the methods we use to generate the synthetic population, in particular the formation model and the probability distributions for the Monte Carlo variables. We then show the resulting planetary formation tracks in the semi-major vs. mass plane. These tracks are of fundamental importance to understand the characteristics of the resulting synthetic population, as they illustrate how and why planets reach their final position in the mass-distance diagram. This $a-M$ distribution at the end of the formation phase is characterized by a number of structures (clumps, concentrations and depletions). Some particular regions in this diagram are identified and discussed. In the companion paper II, we use a synthetic detection bias for the radial velocity method to identify the subset of detectable synthetic planets. We then compare this sub-population with statistical, quantitative methods to the actual extrasolar planet population. In these first two papers, we assume a mass of the host star of 1 $\msun$, as most known extrasolar planets are found around solar type stars. In later papers in this series we shall study the influence of different host star masses as well as of different formation environments (\textit{i.e.} disk properties). Some of these have already been observationally identified such as the well known correlation between the stellar metallicity and the detection probability of giant planets.  

The outline of the paper is as follows: In section \S\ref{sect:giantplanetformationmodel} we give an overview of our formation model, with a focus on the modifications and necessary simplifications\footnote{The synthesis of a population of $\sim30\,000$ planets takes several days on a 50 CPU cluster. Most of the time is spent in solving the planetary envelope structure equations.} relative to Alibert et al. (\cite{alibertetal2005a}). In section \S\ref{sect:montecarlomethod} the Monte Carlo approach is described, whereas we determine the probability distributions for the initial conditions in \S \ref{sect:probabilitydistributions}. As results, section \S\ref{sect:results} illustrates  the numerical population synthesis process with example  formation tracks in the distance-mass plane and discusses the properties of the planet population.  The conclusions are drawn in the last section, \S\ref{sect:conclusions}.

\section{Giant planet formation model}\label{sect:giantplanetformationmodel}
The link between  the initial conditions \textit{i.e.} the properties of the protoplanetary disk, and the final outcome \textit{i.e.} the planets, can only be given by a theoretical formation model.  This link is usually very complicated involving many feed-back mechanisms and nonlinearities.  For the simulations presented in this paper, we calculate in a consistent way the formation of a protoplanet, its migration, and the structure and evolution of the protoplanetary disk. 

\subsection{Disk structure and evolution}\label{subsect:diskstructureandevolution}
The structure and evolution of the protoplanetary disk is modeled as a non-irradiated, 1+1D $\alpha$-disk (Shakura \& Sunyaev \cite{SS}), following the method originally presented in Papaloizou \& Terquem (\cite{PT99}).  
We thus solve the diffusion equation (effective viscosity $\tilde{\nu}$) describing the evolution of the gas surface density $\Sigma$ as a function of time $t$ and distance $a$ to the star:
\beq
{d \Sigma \over d t} = {3 \over a} {\partial \over \partial a } \left[ a^{1/2} 
{\partial \over \partial a} ( \tilde{\nu} \Sigma a^{1/2})  \right] + \dot{\Sigma}_w(a)
\label{eqdiff_std}
\eeq
The photo-evaporation term $\dot{\Sigma}_w$ is given by (Veras \& Armitage \cite{VA04}):
\beq
\dot{\Sigma}_w  =\left\{ \begin{array}{ll}
0 & \textrm{for}\  a < R_g \\
\frac{\mwind}{2 \pi (\amax-R_{g}) a} & \textrm{otherwise}
\end{array} \right.
\eeq
where $R_g$ is taken to be $5$ AU, $\amax$ is the size of the disk, and the total mass loss $\mwind$ due to photo-evaporation is an input parameter which together with the $\alpha$ parameter determines the lifetime of the disk. 

For simplicity, we adopt an initial profile of the gas disk surface density according to the phenomenological model of Hayashi (\cite{hayashi1981}), $\Sigma(a,t=0)=\sigmanorm \left(a/\anorm\right)^{-3/2}$ where $\sigmanorm$ is the surface density at our reference distance ($\anorm$=5.2 AU), and the computational disk extends from $\amin$=0.1 AU to $\amax$=30 AU. The initial total gas disk mass in the computational disk is then $4 \pi \sigmanorm \anorm^{3/2}(\amax^{1/2}-\amin^{1/2})$. For the initial profile with $\Sigma\propto a^{-3/2}$ the accretion rate decreases from the inner to the outer parts of the disk.  As shown in e.g. Papaloizou \& Terquem (\cite{PT99}), the inner parts of the disk evolve rapidly toward a state of constant accretion rate $\mdotst$. Therefore, the inner initial gas disk profile is truncated in order to obtain an accretion rate lower that a constant value of order $3 \times 10^{-7} \msun / $yr. This allows us to speed up the calculation of the disk evolution. 

\begin{figure}
   \resizebox{\hsize}{!}{\includegraphics{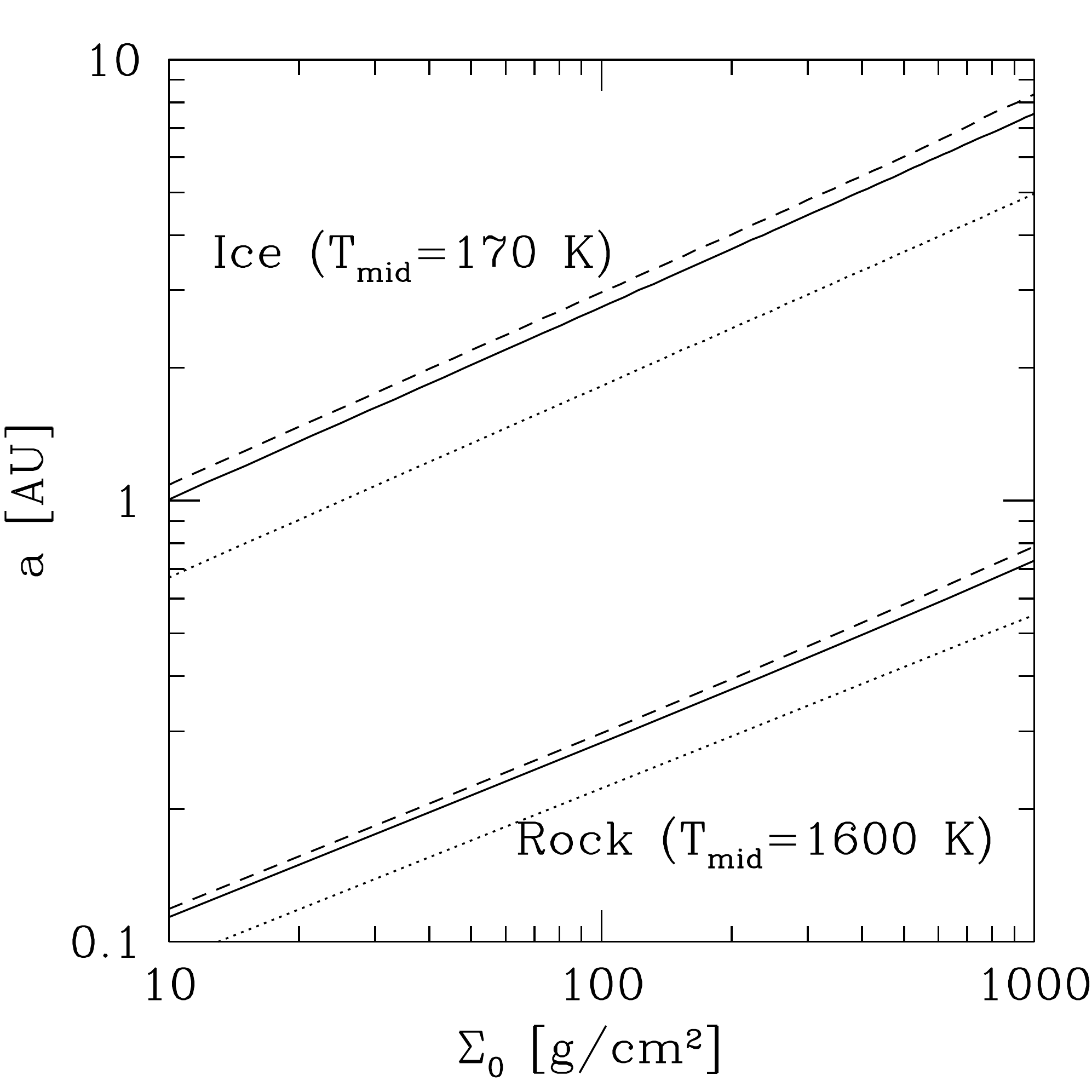}}
   \caption{Position of the iceline $\aice$ as a function of the
    initial gas surface density $\sigmanorm$ at 5.2 AU
   (upper three lines).  It corresponds to initial $T_{\rm mid}$ of
   170 K.  The iceline is plotted for three values of $\alpha$: 0.01 (dashed line), 0.007 (solid line) and 
   0.001 (dotted line). The lower three lines correspond to
   initial $T_{\rm mid}$ of 1600 K, roughly the
   evaporation temperature of rock.  The rockline $\arock$ is however
   not  taken into account in the nominal model, due to the
   difficulty defining its relevant location, as disk evolution is very rapid close-in and irradiation effects might be important  (cf. paper II).} 
   \label{fig:rrockrice}
\end{figure}

The initial solid surface density is given by $\sigmad= \fpg \fri \sigmanorm\left(a/\anorm\right)^{-3/2}$ (Hayashi \cite{hayashi1981}; Weidenschilling et al. \cite{weidenschillingetal1997}) where $\fpg$ is the dust-to-gas ratio of the disk, and $\fri$ is a factor describing the degree of condensation of ices.  Its value is set to $1/4$ in the regions of the disk for which the initial mid-plane temperature exceeds the sublimation of water ice ($T_{\rm mid} > 170$K),  and 1 otherwise.  The semimajor axes $\aice$ where this temperature is reached as a function of initial gas surface density $\sigmanorm$ is plotted in fig. \ref{fig:rrockrice}. For a minimum mass solar nebula (MMSN) like $\sigmanorm$ (100-200 g/cm$^2$), the iceline is as expected found between 2 and 4 AU (Hayashi \cite{hayashi1981}). Note that in the active disk model we use, the effect of stellar irradiation on the temperature structure of the disk is not included. 
  
\subsection{Migration rate}\label{subsect:migrationrate}
The migration of the protoplanet occurs in two main regimes depending upon its mass.  Low mass planets undergo type I migration (Ward \cite{Ward}; Tanaka et al. \cite{Tanaka}) which  depends linearly on the body's mass.  The prevalence of extrasolar planets have led to suspect that the actual type I migration rate is probably significantly lower than  currently estimated (Menou \& Goodman \cite{menou}; Nelson \& Papaloizou \cite{NelsonPap04}). For this reason, we allow for a arbitrary reduction of the type I migration rate as calculated in Tanaka et al. (\cite{Tanaka})  by a constant efficiency factor $\f1$.

The migration type changes from type I to type II when the planet becomes  massive enough to open a gap in the disk. We assume that this happens when the Hills radius of the planet becomes greater than the density scale height $\tilde H$ of the disk (Lin \& Papaloizou \cite{LP86}). Planetary masses were the migration regime changes can be low with such a thermal criterium only, as found also by Papaloizou \& Terquem (\cite{PT99}) who use a similar condition. This is especially the case as due to disk evolution,  the disk scale height  $\tilde H$ decreases with time, so that the minimal mass needed to open a gap decreases. This effect is emphasized by the fact that our disk model does currently not include irradiation, so that especially towards the end of disk evolution, $\tilde H$ gets smaller than in a disk including it, and smaller planets can open a gap (Edgar et al. \cite{edgaretal2007}).  The order of magnitude we obtain is however consistent with the one derived from Armitage \& Rice (\cite{armitagerice2005}), since they give a gap opening  condition (including the effect of viscosity) of $ \mplanet / \mstar \gtrsim \alpha^{1/2}(\tilde H /\aplanet)^2$. In our simulation, the transition typically occurs when the aspect ratio of the disk has become tiny, between 2 and 3\%, meaning a transition at tens of Earth masses. We note that  Crida et al. (\cite{cridaetal2006}) have derived a new criterion for gap opening which depends on both the disk aspect ratio and the Reynolds number. Using such a modified transition mass has some influence on the planetary formation tracks (see sect. \ref{subsubsect:transitiontypeItypeII}).

Type II migration (Ward \cite{Ward}) itself comes in two forms: As long as the local disk mass is  large compared to the planet's mass $\mplanet$ (called ``disk dominated'' migration in Armitage \cite{armitage2007}), the planet is coupled to the viscous evolution of the disk and its migration rate is independent of its mass. The planetary migration timescale is then the same as the gas viscous timescale (e.g. Ida \& Lin \cite{idalin2004a}). Once the local disk mass and the planet's mass become comparable migration slows down (Lin \& Papaloizou \cite{LP86}) and eventually stops. Due to the inertia of the planet the disk can no longer deliver the amount of angular momentum necessary to force the planet to migrate at the gas' radial speed (e.g. Trilling et al. \cite{trillingetal1998}, called ``planet dominated'' migration in Armitage  \cite{armitage2007}).

\begin{figure}
   \resizebox{\hsize}{!}{\includegraphics{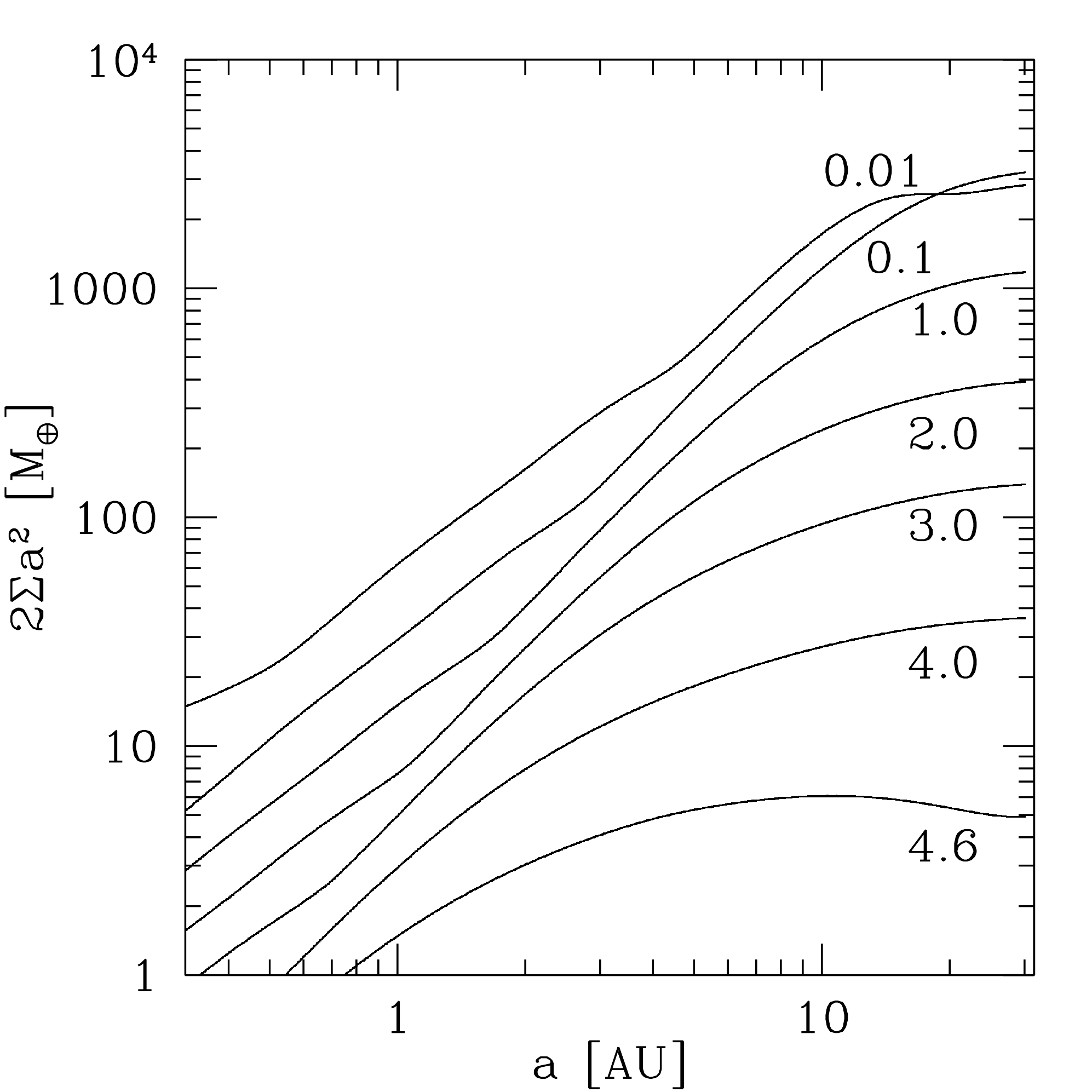}}
   \caption{Example of the evolution of  $2 \Sigma a^2$. Seven different moments in time are plotted (time in units of Myrs).  At 4.6 Myr  the disk has nearly vanished, and the calculations are stopped.} 
   \label{fig:sigmaaalocaldiskmass}
\end{figure}

As Armitage (\cite{armitage2007}) and Thommes et al. (\cite{thommesetal2008}), we have found that this braking phase plays a key role in determining the final semi-major axis of massive planets.  The reason for this can be seen in fig.  \ref{fig:sigmaaalocaldiskmass} where $2 \Sigma a^2$ is plotted as a function of time and semimajor axis for an example disk evolving under  the influence of viscosity and photo-evaporation. The quantity $2 \Sigma a^2$ serves as the measure of the local disk mass to which the planet's mass is compared (Lin \& Papaloizou \cite{LP86}; Syer \& Clarke \cite{syerclarke1995}; Armitage \cite{armitage2007}). 

The plot shows that except at the very end, $2 \Sigma a^2$ always increases  with $a$. Initially, in the region where most giant planet begin their formation ($\sim$5-10 AU), a mass of at least a few Jupiter masses is needed to get into the braking phase. However, after 1-2 Myr,  which is the typical timescale to build protoplanets that have a sufficient mass to migrate in type II, a mass of the order of $\sim 10 \mearth$ at $\sim1$ AU is already sufficient to enter into the slower planet dominated type II mode. Combined with eq. \ref{eq:migrationrateduringbraking}, fig.  \ref{fig:sigmaaalocaldiskmass} also shows that once the braking starts, it steadily increases as  $2\Sigma a^2$ decreases with decreasing $a$ and with time, while the planet's mass can continue to grow. 

A consequence of  the temporal evolution of the gaseous disk is  the slowing down of  type II migration rate thus providing a natural mechanism to halt planets at intermediate distances. At  very small distances from the star ($\lesssim 0.1$ AU),  other, special stopping mechanisms might additionally be at work  (Lin et al. \cite{linetal96}). 

As in Alibert et al. (\cite{alibertetal2005a}), the migration rate during the braking phase  ($\mplanet>2 \Sigma(\aplanet) \aplanet^2$) is  calculated as
\beq\label{eq:migrationrateduringbraking}
\frac{d \aplanet}{dt}=-\frac{3 \nu}{\aplanet}\frac{\Sigma(\aplanet,t) \aplanet^2}{\mplanet},
\eeq
where $\aplanet$ is the semimajor axis and $\nu$ the viscosity. This is a modification of eq. 64 in Ida \& Lin (\cite{idalin2004a}) in the sense that we use as Edgar (2007) the disk properties like $\Sigma$ directly at the planet's position and not at the radius of maximum viscous coupling,  assuming that for the rather high $\alpha$ we are using, wave dissipation and thus angular momentum exchange will occur essentially in the proximity of  the planet (Lin \& Papaloizou \cite{linpapaloizou1984}).  

It can finally also be noted, that a slowing down of $da/dt$ $\propto\Sigma \aplanet^2 / \mplanet$ explains naturally why larger planets should stop further out, provided that $\Sigma a^2$ increases with $a$.  This behavior is indicated by observations (Zucker \& Mazeh \cite{zuckermazeh2002})

\subsection{Protoplanet structure and evolution}\label{subsect:protoplanetstructureandevo}
The structure of the forming planetary envelope is calculated by solving the standard equations of planet evolution as in Alibert et al. (\cite{alibertetal2005a}),
but  assuming that the luminosity of the envelope $L$ is uniform, and equal to the accretion luminosity of  planetesimals: 
\beq	
L ={ G \mcore \mdotcore \over \rcore }.
\eeq
In this equation, $\mdotcore$ is  the accretion rate of planetesimals, $\mcore$ is the mass of the core of the planet,  and $\rcore$ its radius. The accretion rate is calculated according to  Greenzweig \& Lissauer (\cite{GL92}), using the same prescription for the planetesimal random velocities  $v_{\rm disp}$ as in Pollack et al. (\cite{pollacketal1996}). We assume a $\mdotcore$ which is independent of migration, for the reasons given in Ida \& Lin (\cite{idalin2008}). We therefore do not explicitly compare the timescales of planetesimal random velocity excitation with the migration timescale to see whether the protoplanet acts as a ``predator'' or ``shepherd'' (Tanaka \& Ida \cite{tanakaida1999}).  It should in any case be noted that if the type I efficiency factors $\f1$ are understood as a consequence of a ``random walk'' type migration (e.g. Nelson \& Papaloizou \cite{NelsonPap04}), where the single modifications of the orbit occur on a short timescale, then we are in a regime that remains yet to be explored in details (Daisaka et al. \cite{daisakaetal2006}). In the slow, planet dominate type II regime, where planets are massive, accretion of planetesimals is usually no more important, as ejection dominates (sect. \ref{subsubsect:outergroup}). When calculating $\mdotcore$, we take into account the focusing effect of the planetary envelope (Inaba \& Ikoma \cite{inabaikoma2003}). The iterative procedure to get the effective capture radius of the planet $\rcapt$ is the same as in Pollack et al. (\cite{pollacketal1996}). In these calculations the effects of ablation are included (Benvenuto \& Brunini \cite{benvenutobrunini2008}). We have found that ignoring the focussing effect of the envelope leads to a planetary population with similar general properties, but where roughly only half as many giant planets can form.

The assumption that $L$ is uniform and due to planetesimal accretion only constitutes a major difference between the models in this work and the ones of Alibert et al. (\cite{alibertetal2005a}): we do not take into account  the exact location of the energy deposition of infalling planetesimals in the envelope,  nor the energy released by the contraction of the envelope.  Tests have shown that these assumptions do not strongly  affect the formation, as also shown by Rice \& Armitage (\cite{ricearmitage2003}).


Solving the structure equations using the local disk temperature and pressure as external boundary conditions gives us the  gas accretion rate of the planet as well as the critical mass for gas runaway accretion. Note that Miguel \& Brunini (\cite{miguelbrunini2008}) have recently shown that the large uncertainties affecting the constants used in parameterized gas accretion laws as in Ida \& Lin (\cite{idalin2004a}) lead to large variations of the predicted final planetary mass distribution.

The afore-mentioned method is valid as long as the disk can supply enough mass to keep  the outer radius equal to the Hill (or the accretion) radius, {\it i.e.}  if the gas accretion rate deduced from the envelope structure calculation is below the maximum rate at which gas can be delivered by the disk onto the planet.  

This latter quantity can be influenced by the response of the disk on the planet's tides. Indeed, hydrodynamical simulations (Lubow et al. \cite{lubowetal1999};  D'Angelo et al. \cite{dangeloetal2002}) have shown that,  when the planet opens a gap in the disk, the accretion rate of gas is highly reduced.  However, it has also been shown (see Kley \& Dirksen \cite{kleydirksen2006}) that when the mass of the planet becomes of the order of $3-5$ $\mj$ ($\mj$ is the mass of Jupiter $\approx 318$ $\mearth$), the disk-planet system can undergo a dynamic instability, leading to a substantial increase of the accretion rate of gas.  For that reason, we assume in this paper that the planetary gas accretion rate in the disk limited case is simply equal to the accretion rate in the disk, namely 
\beq\label{eq:dmdtdisklimited}
\frac{d\mplanet}{dt} = \dot{M}_{\rm disk} = 3 \pi \tilde{\nu} \Sigma. 
\eeq
This setting constitutes another difference to the models in Alibert et al. (\cite{alibertetal2005a}), where we had limited the planet's accretion rate across a gap according to Veras \& Armitage (\cite{VA04}).

As mentioned in Alibert et al. (\cite{alibertetal2005a}), since we are primarily interested in the mass and semi-major axis evolution of forming planets, the planet internal structure is no more calculated once the limiting accretion rate  $\dot{M}_{\rm disk}$ is reached. Therefore, in this second phase, we can no more explicitly compute the capture radius $\rcapt$ of the planet. Rather, it is simply assumed to scale with the core radius, \textit{i.e.} the ratio $\rcapt/\rcore$ is kept constant. As a consequence, the amount of solids accreted after the limiting accretion rate as been reached is uncertain. This affects however mainly large, gas dominated mass planets, and not low mass ones (like ``Hot Neptunes'').

\subsection{Limitations of the model}\label{subsect:limitationsofthemodel}
From its conception, our model is well suited to describe the formation of giant  gaseous planets, but only to a lesser extend the formation of very low mass (terrestrial) planets. This is mainly due to three assumptions that are made. Additionally, our model is a formation, not an evolutionary model (\S \ref{subsect:limitationshotplanets}).
 
\subsubsection{Initial embryo mass}\label{subsubsect:limitationembryomass}
First, we always assume an initial seed embryo mass of $\membstart$=0.6 $\mearth$, similar to Pollack et al. (\cite{pollacketal1996}) or Bodenheimer \& Pollack (\cite{bodenheimerpollack1986}), as at such a mass the opacity in the envelope becomes sufficient to justify the diffusion approximation for the radiative flux  (Bodenheimer \& Pollack \cite{bodenheimerpollack1986}).  This assumption implies that our models are only valid for planets with final masses exceeding this value by some significant margin. More quantitatively, the assumption of  $\membstart$=0.6 $\mearth$  is reasonable when the local isolation mass  $\miso$ (cf. \S \ref{subsect:embryostarposition}) is significantly larger than 0.6 $\mearth$. For disks  similar to the MMSN,  $\miso$ is larger than 0.6 $\mearth$  only beyond the iceline $\aice$ (e.g. Lissauer \& Stewart \cite{lissauerstewart1993}).

\subsubsection{Growth after disk dispersal}\label{limitationslatetimeevo}
Second, we stop the calculations when the gas disk has disappeared (or when the planet has migrated close to the sun, cf. below), and ignore all processes taking place later on. While for giant gaseous planets this assumption is reasonable (except for effects due to the concurrent growth of many planets, see below), inside the iceline, for disks similar to the MMSN,  growth from $\miso\sim0.01-0.1$ $\mearth$ to the final masses occurs through giant impacts on timescales that exceed by roughly one order of magnitude typical gas disk lifetimes (Goldreich et al. \cite{goldreichetal2004a}). Thus, for terrestrial mass planets, growth after the dispersal of the gas disk is of large importance. Planetary accretion proceeds at a slower pace at larger distances (\S \ref{subsection:embryostarttime}), so that our assumption of an essentially completed formation of the planets at the time of disk dispersion is also not fulfilled for the formation of gas-free ice giants at large distances ($a\gtrsim10-30$ AU, Ida \& Lin \cite{idalin2004a}).

We therefore caution that our synthetic planetary populations are incomplete for masses less than a few earth masses for $a<\aice$ and less than a few 10 $\mearth$ for $a>\aice$. 
We stress however, that the fact that the vast majority of seed embryos do not become giant planets (cf. below) is not an artifact of the model but a consequence of the fact that the most common protoplanetary disks allow only the formation of relatively low mass planets. 

\subsubsection{One embryo per disk approach}
An important limitation of the model that is linked to the last point is the fact that we follow, as in previous similarly detailed giant planet formation models (Pollack et al. \cite{pollacketal1996}; Alibert et al. \cite{alibertetal2005a}), the growth of only one embryo per disk, revolving on a circular orbit. 

In reality it is clear that more than one embryo will emerge in the same protoplanetary disk, which typically start out forming in rapid succession and close proximity (Thommes et al. \cite{thommesetal2008}). This multiplicity can have several kinds of effects during formation, as shown by Thommes et al. (2008): Gravitational interactions between forming planets can modify their migration rate during formation, in particular by locking into resonances. Moreover, similar interactions can lead to modification of the semi-major and eccentricity distributions, also after the formation process itself (see also Adams \& Laughlin \cite{adamslaughlin2003}). In addition, planets forming in the same protoplanetary disk act as competitors for planetesimals and gas accretion, where e.g. one planet can cut off the gas supply of the other ones. The competition for planetesimals accretion was in particular addressed by Alibert et al. (\cite{alibertetal2005b}) in the case of the Solar System formation, and of the \object{HD\,69830} system (Alibert et al. \cite{alibertnewsyst2006}). In these two models, the internal structure of forming planets were calculated (as in the present models), but not the gravitational interactions between forming planets. In a different, and nicely complementary approach, Thommes et al. (\cite{thommesetal2008}) have calculated a large set of multi-planet formation models (following a much larger number of embryos compared to the two afore-mentioned studies but without determining the internal structure of planets) accounting for both competition for gas and solids accretion, and gravitational interactions between planets. Including several embryos while keeping the detailed physics describing one single embryo is a difficult, but important step to be taken in future models.

\subsubsection{Planets very close to the star}\label{subsect:limitationshotplanets}
Other complications arise when planets migrate very close to their host star ($\lesssim 0.1$ AU). This close to the star, the disk structure is more complex than further out due for example to magnetic field effects (Lin et al. \cite{linetal96}) or tidal interactions  (Trilling et al. \cite{trillingetal2002}), which influence the formation of planets entering this zone, by altering the  accretion or the migration rate (Papaloizou \& Terquem \cite{PT99}). Also after formation, very close-in planets can be subject to mass loss by evaporation (Vidal-Madjar et al. \cite{vidalmadjaretal2003}; Baraffe \cite {baraffeetal2004}). Planets in great proximity to the star can thus only be described if a more detailed disk model is used, if additional stopping mechanisms for migration are taken into account and finally, if a subsequent  evolutionary model for evaporation (Baraffe et al. \cite{baraffeetal2006}) is included. These complications also illustrate the importance of discovering more (small) planets at ``safe''  distances from the star, as they are a more direct constraint on formation models.

Our model currently doesn't include any of the afore-mentioned effects. We therefore simply stop the calculations when a planet of mass $\mplanet$ has migrated to $\atouch$, which is defined as the semimajor axis where the inner boundary of the planet's feeding zone touches the inner boundary of our computational disk at $\amin$=0.1 AU, (``the feeding limit'') \textit{i.e.} at
$\atouch=\amin/(1-4(\mplanet/(3\mstar))^{1/3})$. If a planet has migrated to $\atouch$, all we can state is that its final semimajor axis would be $\leq\atouch$ (it is also possible that it eventually would have fallen into the host star), and what its mass at $\atouch$ was.
  
\section{Monte Carlo method}\label{sect:montecarlomethod}
The basic idea of using a Monte Carlo method to synthesize planetary populations is to sample all possible combinations of initial conditions (protoplanetary disk mass, metallicity, etc.) with a realistic probability of occurrence. This leads to all possible final outcomes of the formation process (\textit{i.e.} planets) also occurring with their relative probabilities.  We first explain the general six steps procedure that we used. 

In the first step, we have identified four crucial initial conditions, and studied the domain of possible values they can take (\S \ref{subsection:variables}). Some other initial conditions had to be kept constant during the synthesis of one population, for simplicity or computational time restrictions (\S \ref{subsect:parameters}). In the second step, we have derived probability distributions for each of the four Monte Carlo variables (\S \ref{sect:probabilitydistributions}). In the third step, we draw in a Monte Carlo fashion large numbers of sets of initial conditions. The forth step consists in using the formation model for each set of initial conditions, giving the temporal evolution of the planet (formation tracks, \S \ref{subsect:exampleevopathes}) as well as its final properties (mass, semimajor axis, composition etc., \S \ref {subsect:masssemimajoraxisdiagramm}).

Many of these synthetic planets would remain undetected by current observational techniques. So, to be able to compare the synthetic  planet population with the observed one, we apply in the fifth step a detailed synthetic detection bias (paper II). In  this way, we obtain the sub-population of \textit{observable} synthetic planets. Ultimately, in the sixth step, we have performed quantitative statistical tests (paper II)  to compare the properties of this observable  synthetic exoplanet sub-population with a comparison sample of real extrasolar planets.

\subsection{Monte Carlo Variables}\label{subsection:variables}
We use four Monte Carlo variables to describe the varying initial conditions for the planetary formation process. Three describe the protoplanetary disk and one the seed embryo.

1. The dust-to-gas ratio in the protoplanetary disk $\fpg$ determines (together with $\sigmanorm$) the solid surface density. Models with $\fpg$ between 0.013 and 0.13  were computed. Combined with the domain of $\sigmanorm$,  this corresponds to initial solid surface densities at $\anorm=5.2$ AU between 0.65 and 130 g/cm$^2$. For comparison, the MMSN has a value of approximately 2.5 g/cm$^2$ (Hayashi \cite{hayashi1981}).

2. The initial gas surface density $\sigmanorm$ at 5.2 AU gives the amount of gas available. Values between between 50 and 1000 g/cm$^2$ were used. The MMSN is estimated to have had a value of about 100-200 g/cm$^2$ (Hayashi \cite{hayashi1981}).

3. The last variable that characterizes a disk is the rate at which it looses mass due to photoevaporation $\mwind$. For the population presented below, it was allowed to vary between $5\times 10^{-10} \msun$/yr and $3 \times 10^{-8}\msun$/yr. 

4.  Finally, the initial semimajor axis of the seed embryo within the disk,  $\astart$, is the fourth variable.  It can take values of 0.1 $\leq \astart \leq$20 AU.

\subsection{Parameters}\label{subsect:parameters}
Some other initial conditions of the model were kept constant for all planets of given population.  We mention only the most important parameters here. More details can be found in Alibert et al. (\cite{alibertetal2005a}). For the nominal population discussed in \S\ref{sect:results}, we use a viscosity parameter $\alpha$ for the disk model of 0.007 and an efficiency factor for type I migration $\f1$ of 0.001. The influence of these two important parameters is briefly discussed in \S \ref{subsubsect:f1variation}, and will be further considered in forthcoming publications. In this and the companion paper the mass of the central star  $\mstar$ is kept constant at 1 $\msun$. 

\section{Probability distributions}\label{sect:probabilitydistributions}
In the next step we determine the probability of occurrence of a certain combination of initial conditions.  In the ideal case, the probability distributions for all our variables would be derived directly from  observations. Unfortunately,  in reality, this is not possible either because in some cases observations do not exist or,  even if they exist, a certain amount of modeling is necessary to to extract the distributions from the observations. 

\subsection{Dust to gas ratio $\fpg$ - [Fe/H]}\label{subsect:dusttogasration}
To establish a link between the dust-to-gas ratio $\fpg$ which is the computational variable required by our model and the corresponding observable,  the stellar metallicity [Fe/H], we assume: (1) the stellar content in heavy element is a good measure of the overall abundance of heavy elements in the disk during formation time. Support for this assumption comes from  the small differences between solar photospheric and meteoritic abundances  (Asplund et al. \cite{asplundetal2005}),  (2) a scaled solar composition and (3) a negligibly small influence of the change of the relative heavy element content on the relative hydrogen content in the comparatively small [Fe/H] domain of interest for planet formation in the solar neighborhood ($-0.5\leq$[Fe/H]$\leq0.5$). Then, similar to Murray et al. (\cite{murray2001}), we can write
\beq\label{eq:fpgofFeH}
\frac{\fpg}{f_{\rm D/G,\odot}}=10^{[\rm Fe/H]}
\eeq
where $f_{\rm D/G,\odot}$ is the dust to gas ratio corresponding to [Fe/H]=0. This formula implies that we assume that iron is a good tracer for the relevant  overall amount of solids available for planet formation. Robinson et al. (\cite{robinsonetal2006}) have found that at given iron abundance, planet host stars are enriched in silicon and nickel over stars without planets, indicating that the above relation is a simplification. 

Measurements of the heavy element abundance in the Sun yield the amount (for complete condensation) of high Z material that existed initially in the form of uniformly mixed fine dust grains. However, what is relevant for our simulations is the concentration of solids in the innermost 20 AU of the disk at a later stage, namely when the dust has evolved into the 100 km planetesimals used in our model. 

As has been shown by Kornet et al. (\cite{kornetetal2001}), the transition from the very early dust phase to the later planetesimal phase involves a number of coupled mechanism of dust-dust and dust-gas interactions like dust settling to the midplane, dust growth by coagulation and radial drift. This leads to a redistribution of the solids within the disk, which can in turn have important effects on planetary formation (Kornet et al. \cite{kornetetal2005}). The key point is  (Kornet et al. \cite{kornetetal2004}) that these processes lead to an increase of the solid to gas ratio in the inner ($\lesssim 10 - 20 $ AU) planet forming regions of the disk by advection of solids from the outer parts where a lot of mass resides. The factor of increase from the initial ``(dust-)''$\fpg$ to the ``(planetesimal-)''$\fpg$ varies depending on the initial conditions and is not completely uniform across  the inner disk, but Kornet et al. (\cite{kornetetal2004}) typically find values of 2 to 4.

To take this effect into account at least to first order, we set the effective planetesimal $\fpg$ in the inner disk to a value about 3 times higher than the value that is inferred from the solar photosphere. Unfortunately, the latter one has been debated recently: Anders \& Grevesse (\cite{andersgrevesse1989}) found a value for the protosolar Z of 0.0189, whereas a more recent study (Lodders \cite{lodders2003}) indicate lower values ($Z$=0.0149). Multiplying this value by roughly 3, gives a $f_{\rm D/G,\odot}$ of $\approx0.04$, the value that we regard as nominal in our simulations. Note that if dust redistribution mechanisms are at work with similar consequences in all disks, we can still use the observed [Fe/H] distribution to scale $\fpg$ for other stars. 

Next, we determine the probability distribution for [Fe/H]. The distribution of the metallicity of solar like stars in the solar neighborhood has been well studied (e.g. Nordstr\"om et al. \cite{nordstrom2004}),  and can be well approximated by a Gaussian distribution (mean $\mu$ and dispersion $\sigma$), \textit{i.e.}  
\beq\label{eq:gaussiandist}
p([Fe/H])=\frac{1}{\sigma \sqrt{2 \pi}} e^{-\frac{([Fe/H]-\mu)^2}{2\sigma^2}}
\eeq
The probability density for $\fpg$ is then found by using eq. \ref{eq:fpgofFeH} and the  fundamental transformation law of probabilities, $|p([Fe/H]) d[Fe/H]|=|p(\fpg)d\fpg|$ leading to 
\beq\label{eq:propdistfpg}
p(\fpg)=\frac{\log(e)}{\fpg \sigma \sqrt{2 \pi} } e^{-\frac{(\log(\fpg)-\tilde{\mu})^2}{2\sigma^2}}
\eeq
where $\tilde{\mu}=\log(f_{\rm D/G,\odot})+\mu$. This is  simply a lognormal distribution. 

\begin{figure}
   \resizebox{\hsize}{!}{\includegraphics{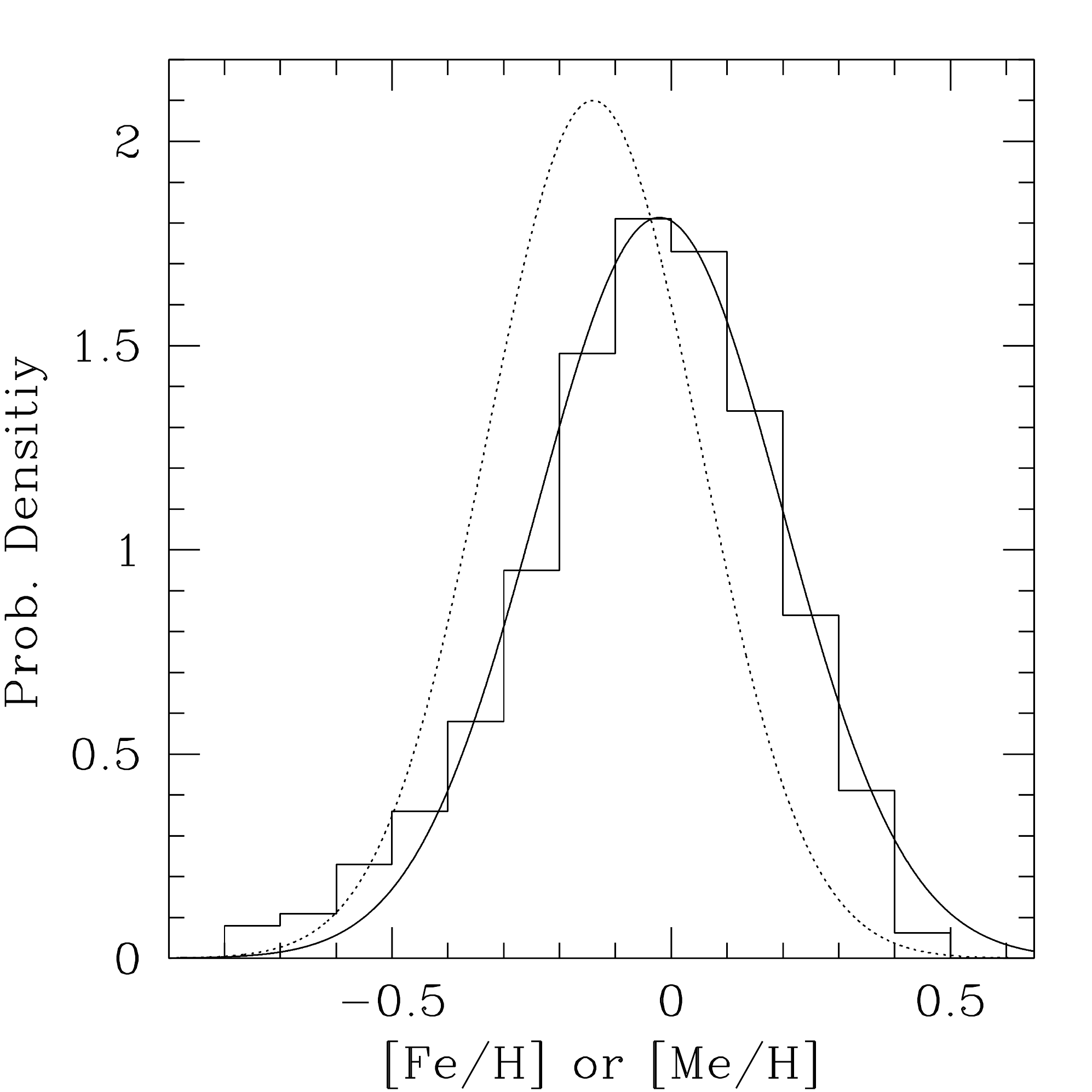}}
   \caption{Solid lines: Histogram and Gaussian fit to the [Fe/H] distribution
   of the CORALIE  planet search sample. Dotted line: For comparison, distribution of
   photometric metallicities  [Me/H] of Nordstr\"om et
   al. (\cite{nordstrom2004}). Probability  densities are given, \textit{i.e.} the
   area under the curves has been normalized to unity in each case.} 
   \label{fig:initialfehdist}
\end{figure}
\begin{table}
\begin{center}
\caption{Parameters $\mu$ and  $\sigma$ describing the Gaussian distribution of [Fe/H],  for different observation samples. FV05 stands for Fischer \& Valenti (\cite{fischervalenti2005}).}
\label{tab:fehparams}
\begin{tabular}{lcc}
\hline\hline
Source& $\mu$   & $\sigma $\\ \hline
Nordstr\"om et al. (\cite{nordstrom2004})& -0.14 &  0.19 \\
Fit to CORALIE planet search sample & -0.02 & 0.22  \\
Fit to FV05 planet search sample &0.05& 0.21\\
Fit to FV05 volume limited sub-sample &-0.05& 0.26\\
\hline
\end{tabular}
\end{center}
\end{table}

Since we aim to quantitatively compare observations and theoretical calculations and compare, for example, real and theoretical detection probabilities, we need the metallicity distribution of a sample of stars that are actually included in planet searches.  Such samples can have a different metallicity distribution than volume limited samples as shown by Fischer \& Valenti (\cite{fischervalenti2005}) who find that their planet search sample is shifted by $\sim0.1$ dex towards higher metallicities relative  to a volume limited subset. The CORALIE planet search sample  (Udry et al. \cite{udryetal2000}) consists of about 1650 G and K dwarf within 50 pc, and is volume limited. In fig. \ref{fig:initialfehdist}  the [Fe/H] distribution (obtained from calibrations to  the CORALIE cross-correlation function) for  about 1000 non-binary, slow rotating stars within the CORALIE search sample  is plotted (Santos et. \cite{santosetal2003}). 

To derive an analytical [Fe/H] probability distribution, we have assumed that the CORALIE data is Gaussian, and performed a non-linear least square fit to obtain the parameters describing this distribution, $\mu$ and $\sigma$.  The results are $\mu=-0.02\approx0.0$ and $\sigma=0.22$ (Table \ref{tab:fehparams}) which we have used in the calculations.  The fit is also plotted in fig. \ref{fig:initialfehdist}.  It approximates the data very well between -0.45$\leq$ [Fe/H]$\leq$0.4.  We have repeated the same procedure also for the planet search sample, and the volume limited sub-sample described in Fischer \& Valenti (\cite{fischervalenti2005}). The results are also given in table \ref{tab:fehparams}. For these two data sets, the Gaussian fit is however somewhat less good, especially for the volume limited sample.

The distribution derived from the CORALIE survey combined with our fiducial value for $f_{\rm D/G,\odot}$,  and the range of $\fpg$, result in the following range of stellar metallicities: $-0.49<$[Fe/H]$<0.51$. This range includes almost all known planet hosting main sequence stars in the solar neighborhood. Finally, we note that despite the complications of linking $\fpg$ to [Fe/H] mentioned here, one can reasonably assume that it is the variable that is most directly constrained by  observational data. 

\subsection{Gas surface density $\sigmanorm$ - Disk Mass $\mdisk$}\label{subsect:gassurfacedensity} 
The probability distribution of gas surface densities $\sigmanorm$ can also be at least partially inferred from observations of star forming regions. The flux density of thermal continuum emission originating from cold dust orbiting young stellar  objects (YSO) at millimeter and submillimeter wavelengths allows  an estimate of the total dust disk mass (Beckwith et al. \cite{beckwith1990}).  By adding  a gas content typical for the interstellar medium  (Beckwith \& Sargent \cite {beckwith1996};  Andrews \& Williams \cite{andrewswilliams2005}), total disk masses  $\mdisk$ are found. The study of many YSO forming concurrently in a given star  formation region then gives distributions for $\mdisk$. 

While the details of these distributions differ across known star formation regions,  most have roughly the shape of a lognormal distribution for $\mdisk$, \textit{i.e.} $\log(\mdisk$)  is Gaussian distributed with a mean $\mu$ and a standard  deviation $\sigma$ (Andrews \& Williams \cite{andrewswilliams2005}; Robinson et al. \cite{robinsonetal2006}). To compute the parameters $\mu$ and $\sigma$ we use Beckwith \& Sargent (\cite {beckwith1996}), who give in their figure 5 histograms for  the distribution of total disk gas masses in the Taurus-Auriga and Ophiuchus star formation region,  and perform as for [Fe/H] a nonlinear least square fit to obtain the two  sets of parameters. 

\begin{figure}
   \resizebox{\hsize}{!}{\includegraphics{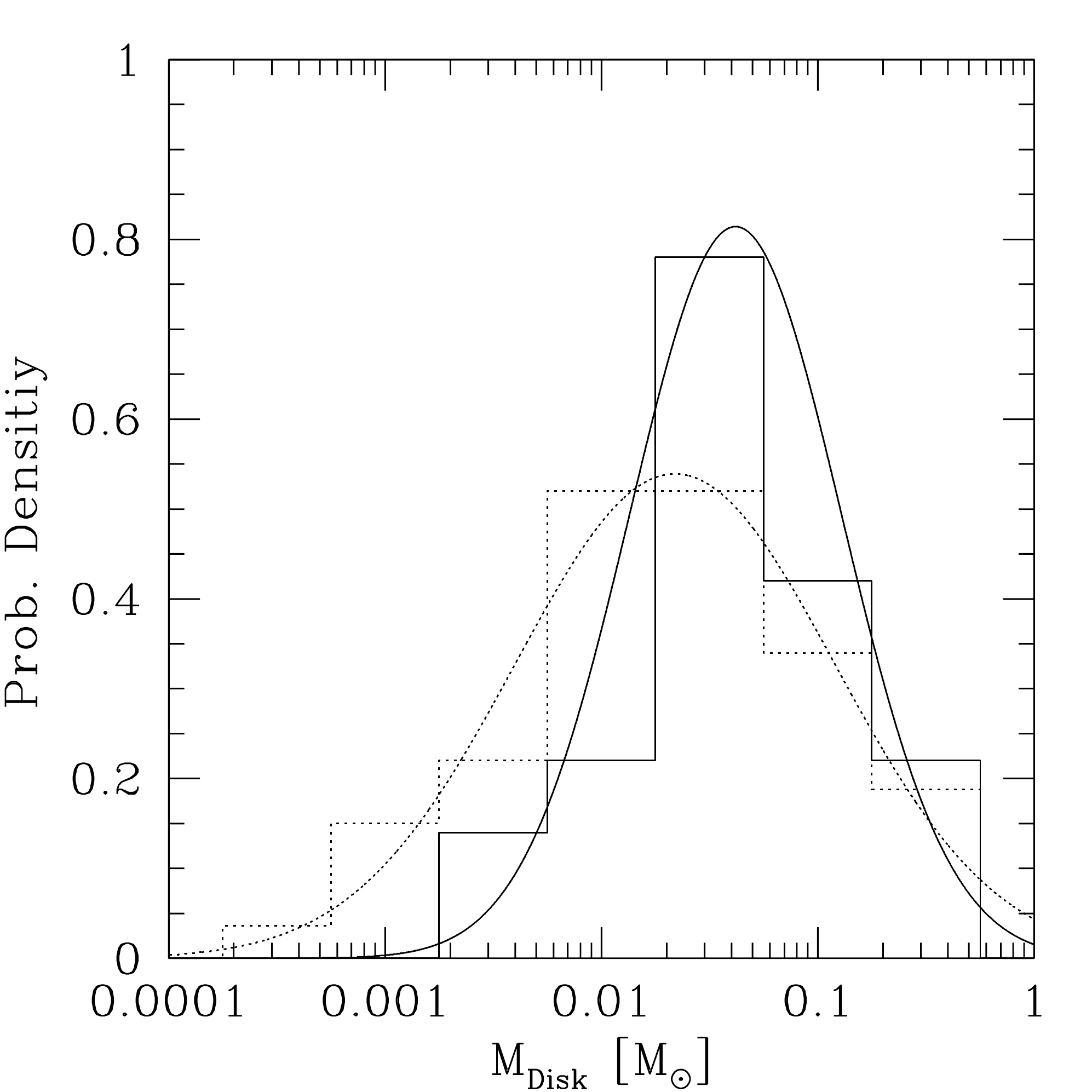}}
   \caption{Histograms and fits to circumstellar disk masses $\mdisk$ for Ophiuchus (solid lines) and Taurus-Auriga (dotted lines). The actual gas masses in the computational domain out to 30 AU are between 0.004 and 0.09 $\msun$.} 
   \label{fig:initialdiskmassdist}
\end{figure}
\begin{table}
\begin{center}
 \caption{Parameters describing the lognormal distribution of
 circumstellar disk masses $\mu$ and $\sigma$, from different sources.
 The value given by Robinson et al. (\cite{robinsonetal2006}) is
 obtained by extrapolating the observed $\mu=-2.31$ of Andrews \&
 Williams (\cite{andrewswilliams2005}) back to the assumed initial state, and
 $\sigma$ is reduced to 0.25 from observed value 0.5. The values of
 Ida \& Lin (\cite{idalin2004a}) are calculated using the value for
 the MMSN from Hayayshi (\cite{hayashi1981}).}\label{tab:Mdiskparams}
\begin{tabular}{lccc}\hline\hline
Source& $\mu$ & $\mdisk(\mu)[\msun]$ &$\sigma$\\ \hline
Fit to Taurus  & -1.66 & 0.022 & 0.74 \\
Fit to Ophiuchus & -1.38 & 0.042 & 0.49\\
Robinson et al. (\cite{robinsonetal2006}) & -1.3 & 0.05 & 0.25 \\
Ida \& Lin (\cite{idalin2004a}) & -1.48 & 0.033 & 1.0 \\
\hline
\end{tabular}
\end{center}
\end{table} 
The observational data, as well as the fits are plotted in fig. \ref{fig:initialdiskmassdist}, while the values  for $\mu$ and $\sigma$ are given in table \ref{tab:Mdiskparams}. For both star formation regions, the disk mass distribution can be reasonably well fitted by Gaussians. The resulting distributions are however very broad which leads to significant probabilities for  disks with masses in excess of 0.1$\msun$ (or even 0.3 $\msun$). For typical thermodynamical conditions, such disks are self-gravitationally unstable in the outer regions (e.g. Ida \& Lin \cite{idalin2004a}). It is likely that this result is due to the presence of the remains of the envelope from which the star formed, contributing to the observed flux (Andrews \& Williams \cite{andrewswilliams2005}).  

The large spread in masses  is also due to the spread in age of the observed YSOs in one cluster (Robinson et al. \cite{robinsonetal2006}) and thus an evolutionary effect. Possibly,  the spread is additionally enhanced by the spread of stellar masses, if the the mass of stars and disks are correlated.

Since our planet formation model includes disk evolution, the mass distribution we require is one at a moment in time as early as possible, before strong evolutionary effects have taken place. Therefore the distribution of Ophiuchus, which is about 2-3 times younger than Taurus-Auriga (White \& Hillenbrand \cite{whitehillenbrand2004}), is more appropriate for our goal and used as nominal distribution for the simulations.

When converting the observed quantity (disk masses), to the corresponding numerical quantity in the model, (the initial gas surface density $\sigmanorm$), we are confronted with a situation similar to the [Fe/H] to $\fpg$ conversion (Matsuo et al. \cite{matsuoetal2007}): Observed protoplanetary disks have physical radii $\aphys$ of typically a few hundred AU (e.g. Beckwith \& Sargent  \cite{beckwith1996}).  For computational reasons we however simulate only the inner part of the disk out to $\amax=30$ AU. Putting the observed disk masses into such a small disk would lead to unrealistically high surface densities. Therefore, when converting the observed disk masses to $\sigmanorm$, we assume that the mass is contained in a disk with a physical radius $\aphys=300$ AU of which we simulate only the innermost 30 AU. The outcomes of our simulations are not very sensitive to the exact value of the assumed physical outer radius, as the disk mass scales only with $\sqrt{\aphys}$. Thus,  the probability density for $\sigmanorm$ is
\beq\label{eq:probabilitydistsigma}
p(\sigmanorm)=\frac{\log(e)}{\sigmanorm \sigma \sqrt{2 \pi} } e^{-\frac{(\log(\sigmanorm)-\tilde{\mu})^2}{2\sigma^2}}
\eeq
where $\tilde{\mu}=\mu-\log\left(4 \pi
\anorm^{3/2}(\sqrt{\aphys}-\sqrt{\amin} \,)\right)$,  which is of a lognormal type (Bronstein et al. \cite{bronsteinsemendjajew1999}).

With the boundaries of the computational disk at $\amin=0.1$ AU and $\amax=30$ AU and given the range of $\sigmanorm$ of 50-1000 g/cm$^2$, we cover a range of disk gas masses in the computational domain between 0.004 and 0.09 $\msun$. These masses should be stable against  self-gravitional collapse (Mayer et al. \cite{mayeretal2004}). 

\subsection{Photo-evaporation rate $\mwind$ - Disk lifetime $\tdisk$} \label{subsection:photoevaporationrate}
The values for the photo-evaporation rate $\mwind$ determines,  together with the viscosity parameter $\alpha$, the timescale $\tdisk$ on which the gas disk disappears. Constraints on its possible values can therefore also be inferred from observation. 

Haisch et al. (\cite{haischetal2001}) have shown by near-infrared observations of hot dust that the circumstellar disk fraction in young stellar clusters is a roughly linearly decreasing function of age with essentially all disk having disappeared after $\sim$6 Myr. They argue that this timescale should be a tracer of the evolution of the bulk of the disk material as well, especially also of the gas disk. A very similar time dependence is also found by Hillenbrand (\cite{hillenbrand2005}).

Assuming a uniform distribution in log for $\mwind$, we have determined the bounds of the $\mwind$ distribution by requiring that the lifetime of our synthetic disks for a fixed value of the viscosity parameter $\alpha$ and for the Ophiuchus distribution of initial disk masses, follows the observed disk lifetime distribution found by Haisch et al. (\cite{haischetal2001}). To determine the bounds, we have first generated a distribution of disk masses as described in sect. \ref{subsect:gassurfacedensity}. We have then set some trial boundaries for the distribution of $\mwind$ and drawn values uniformly in log inside these limits. We can then calculate the resulting lifetime  $\tdisk(\alpha, \sigmanorm, \mwind)$ for each synthetic disk, so that we get a distribution of synthetic $\tdisk$. The bounds were then adjusted in an iterative fashion until the observed and the synthetic disk lifetime distribution are similar.

It is found that a reasonable fit is obtained by allowing a range between $5 \times10^{-10}$  and  $3 \times 10^{-8}$ M$_\odot$/yr for $\alpha=7\times 10^{-3}$. For other $\alpha$, the same iterative procedure was repeated. As one expects, the higher $\alpha$, the lower the necessary $\mwind$ boundaries in order to reproduce Haisch et al. (\cite{haischetal2001}). We have for example found that for $\alpha=10^{-2}$, bounds equal   $1 \times10^{-10}$  and  $1.5 \times 10^{-8}$ M$_\odot$/yr are appropriate. Such values are compatible to the ones found elsewhere (Armitage et al. \cite{armitageetal2003}). 

\begin{figure}
   \resizebox{\hsize}{!}{\includegraphics{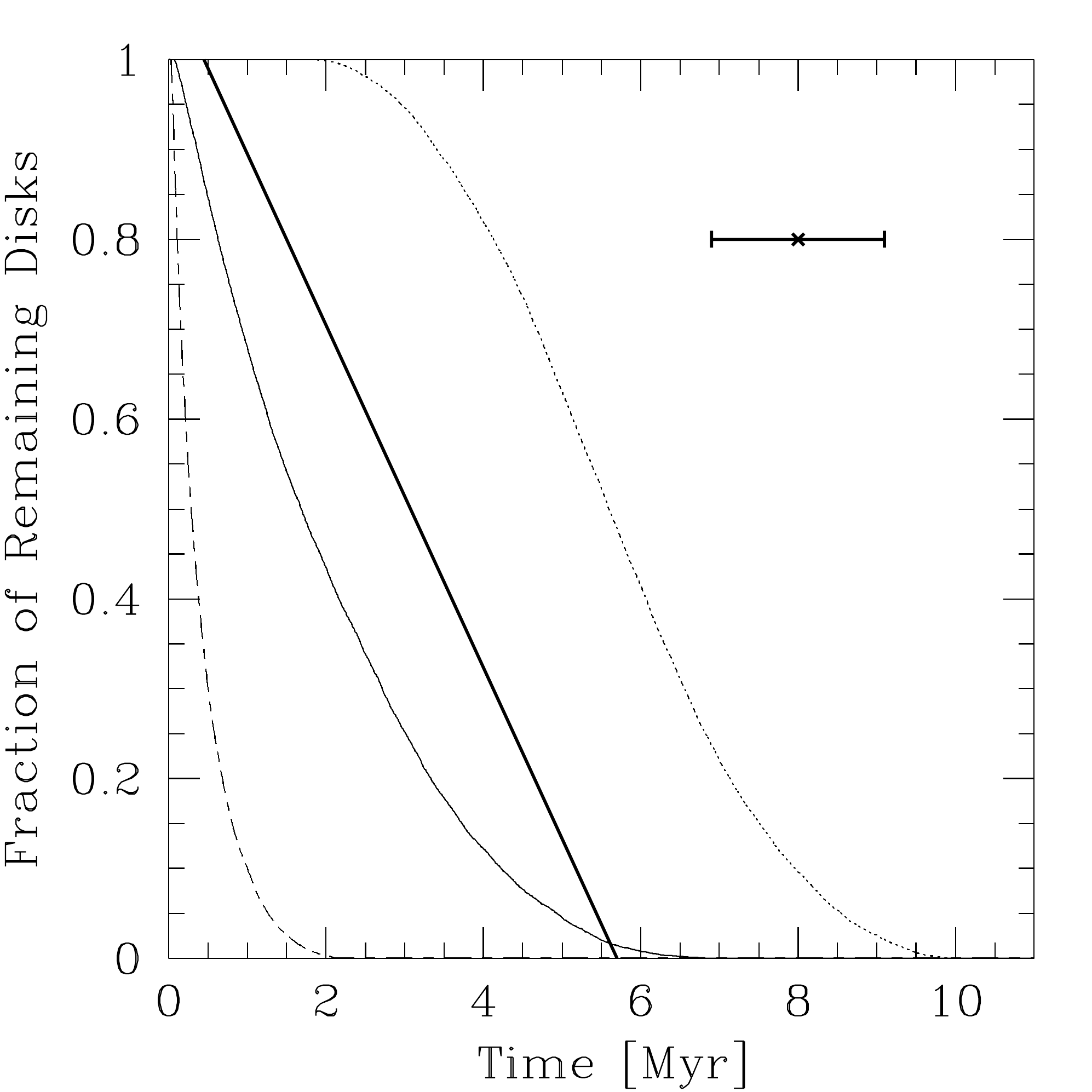}}
   \caption{Fraction of stars in the model possessing a gaseous disk as a function
   of time for a uniform distribution in log of $\mwind$ between $ 5
   \times 10^{-10}$ and $3 \times10^{-8}$ M$_\odot$/yr (thin
   solid line). The thick solid line is the fit of Haisch et
   al. (\cite{haischetal2001}) to the  observed JHKL excess / disk
   fraction as a function of mean cluster age. The error bar at the
   top right is the overall systematic uncertainty in age of these
   observations, also from Haisch et al. (\cite{haischetal2001}). For comparison, the
   fraction is also given if two different distributions of $\mwind$ are used in the model:  Uniformly
   in log between $1 \times10^{-8}-1\times10^{-7} \msun$/yr
   (dash-dotted line),  leading to  disk lifetimes that are clearly too short, or uniform
   in log in $1\times10^{-10}-1\times10^{-9}  \msun$/yr, leading to
   disk lifetimes that are too long (dotted line).} \label{fig:remainingdiskfrac}
\end{figure}

In fig. \ref{fig:remainingdiskfrac} the fraction of stars in the model with remaining disks using the nominal $\mwind$ distribution and two comparison cases are plotted, together with the fit of Haisch et al. (\cite{haischetal2001}). The model and the observed distribution have a similar decrease of the remaining disk fraction if the systematic errors of the ages in the observational data are taken into account. 

It is clear that the distribution of $\mwind$ could in reality be much more complicated and that  also the assumption of a temporally constant UV flux might not be justified. One should for example, when external UV sources are considered,  take into account the stellar environment as O or B stars in rich clusters can greatly enhance the far-ultraviolet flux (Adams et al. \cite{adamsetal2004}; Armitage \cite{armitage2000}). 

The boundary conditions  of our disk model (\S \ref{subsect:diskstructureandevolution}) are similar to the ones presented in Alibert et al. (\cite{alibertetal2005a}). In particular, at the outer disk boundary, we assume no mass influx from the outer parts of the disk. This assumption could lead to an underestimation of the disk dispersal time, since the outer parts of the disk could act as a mass reservoir. However, as explained above, we adjust the photoevaporation rate in order to obtain disk lifetime similar, by construction, to observed ones. With the initial power law surface density, the disk accretion rate is found to decreased towards the outer parts of the disk, reaching values below $10^{-9} \msun$/yr, which is of the order of, or lower than, the photoevaporation rate. Therefore, a large part of the mass present in the outer disk parts is photoevaporated rather than accreted into the inner parts. This means that if we would include inflow, we would have to increase the required photoevaporation rates to obtain disk lifetimes consistent with observation, but otherwise our results would remain similar. 

\subsection{Embryo start position $\astart$}\label{subsect:embryostarposition}
The starting position of the planetary embryos $\astart$ inside the disk cannot be constrained by observations. Therefore, we derive a probability distribution using theoretical arguments only.

The derivation can be made relatively easily if one assumes that during the early phases of planetary accretion radial motions  can be neglected. In this case, it follows (e.g. Lissauer \& Stewart \cite{lissauerstewart1993}) from the restricted 3-body problem that an embryo can accrete background planetesimals only within its feeding zone which has a half width of $B_L$ times ($B_L\approx3-5$) its Hill sphere radius $\rhill$.  The Hill sphere radius of the planetary embryo at a semimajor axis $a$ and a mass $\membstart$ is given by $\rhill=\left(\membstart/(3 \mstar0\right))^{1/3} a$, which is, for fixed $\membstart$, $\propto$ $a$. Thus, as confirmed by many different numerical simulations (e.g. Weidenschilling et al. \cite{weidenschillingetal1997}; Kokubo \& Ida \cite{kokuboida2000}) runaway bodies emerge with relative separations of their semimajor axis $\Delta$  proportional to their semimajor axes $a$, so $\Delta /a=$ const. This situation prevails also later during oligarchic growth stages, although the numerical value of $B_L$ increases  (Ida \& Lin \cite{idalin2004a}). 

The probability $p(a)$ of starting in an interval $da$ is inversely proportional to the spacing $\Delta$, so    
\beq
p(a) da \propto \frac{da}{\Delta} \propto \frac{da}{a} = d \log (a) \propto {\rm const.},
\eeq
which means that the probability distribution for the starting position is uniform in log, as assumed also by Ida \& Lin (\cite{idalin2004a}).

While the distribution of the initial position is clear, there remains the issue that we start our simulations with a seed embryo of mass $\membstart=0.6$ $\mearth$.  In order to be self-consistent, we have to require that this amount of mass is available at the starting position. In other word, we have to ensure that for a set of initial values  ($\fpg$, $\sigmanorm$ and $\astart$) there is indeed enough mass available to build-up such a seed. For this, we consider the amount of heavy elements in the planet's feeding zone $2\pi\astart 2 B_L \rhill \sigmad$ to calculate the isolation mass $\miso$ (Lissauer \& Stewart \cite{lissauerstewart1993}):   
\beq\label{eq:isomassgeneral}
\miso=  \frac{(4 \pi B_L \astart^2 \sigmad)^{3/2}}{(3 \mstar)^{1/2}}
\eeq 
For the disk profile we use this becomes
\beq\label{eq:isomassspec}
\miso=\frac{(4 \pi B_L \fpg \fri \sigmanorm \anorm^{3/2} \astart^{1/2})^{3/2}}{(3 \mstar)^{1/2}}, 
\eeq
where $\fri$ is itself a function of $\sigmanorm$ (and $\alpha$, fig. \ref{fig:rrockrice}). 

If the isolation mass calculated from eq. \ref{eq:isomassspec} with $B_L=2\sqrt{3}$ (Lissauer \& Stewart \cite{lissauerstewart1993}) for a set of  $\fpg$, $\sigmanorm$ and $\astart$  exceeds 0.6 $\mearth$, we conclude that these are self-consistent  initial conditions for the formation model. 

As $\miso$ $\propto$ $\astart^{3/4}$, the isolation mass criterion sets an inner boarder for the possible domain  of $\astart$. We also require that the starting time of the embryo $\tstart$ is smaller than the disk lifetime $\tdisk$ (cf. the next section \S\ref{subsection:embryostarttime}). In practice, we first calculate for a given  $\fpg$, $\sigmanorm$ and $\mwind$ the range of $\astart$ in which $\miso\geq\membstart$ and $\tdisk\geq\tstart$,  and then draw $\astart$ from that range. 

\begin{figure}
   \resizebox{\hsize}{!}{\includegraphics{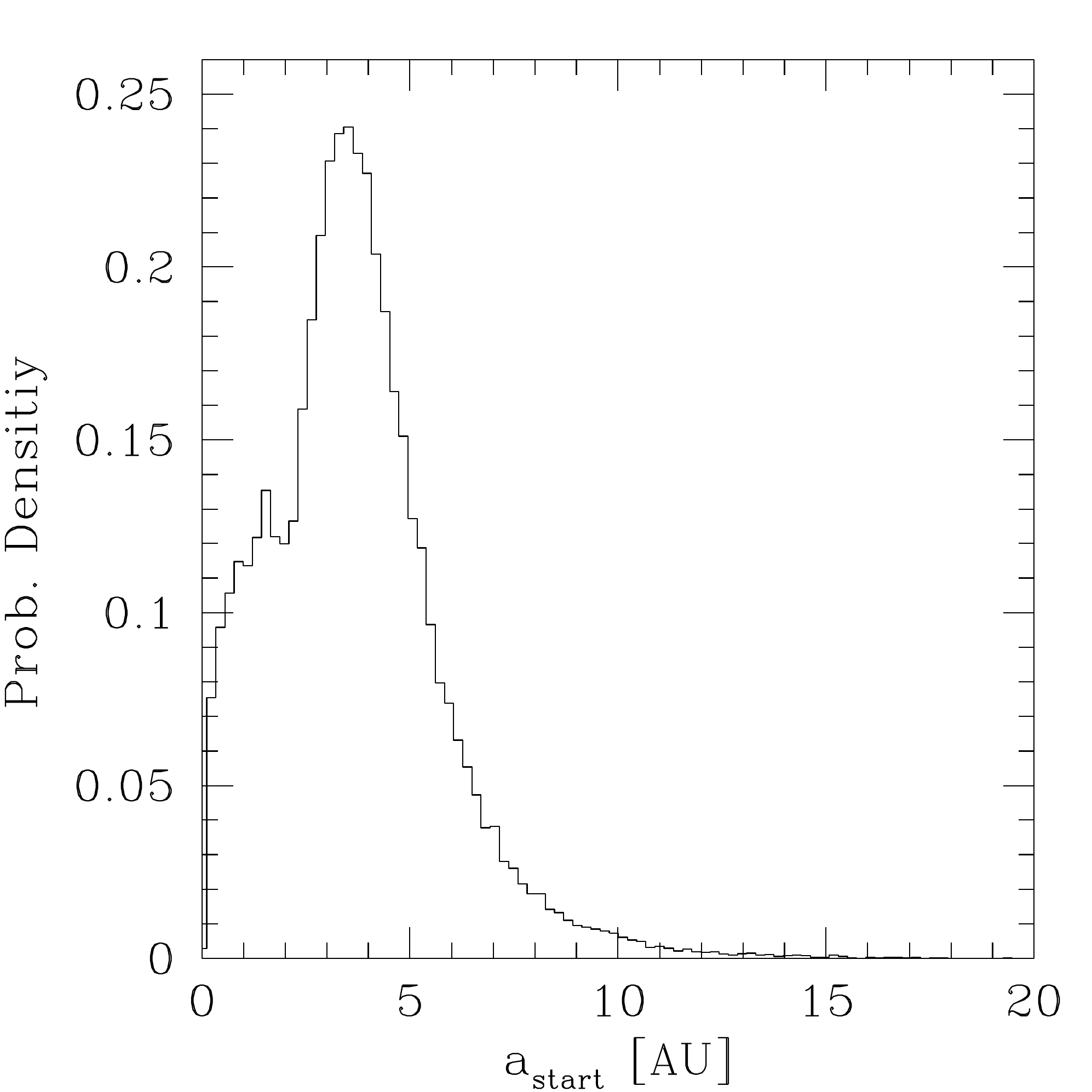}}
   \caption{Probability distribution for $\astart$. There is a marked
   peak around 3 AU. Outside this distance,  the probability
   distribution falls off as expected for a distribution that is uniform in
   log. The fraction inside about 2 AU  is due to disks with the ability
   to form seeds sufficiently massive  ($\geq \membstart$) inside the iceline
   as well.} 
   \label{fig:a0dist}
\end{figure}

In fig. \ref{fig:a0dist} the resulting distribution of $\astart$ is plotted.  Its particular shape is the results of the combination of the probability distribution for  $\astart$ (uniform in log) and the criteria that $\tdisk\geq\tstart$ and $\miso\geq\membstart$.  The fact that the latter is fulfilled only beyond the iceline for most disks,  and the property of the uniform-in-log distribution to emphasize  the smallest possible values lead to the peak near 3 AU. Further out,  the probability decreases, a consequence of the uniform-in-log  distribution and the increasing $\tstart$. Only disks with a high surface density (\textit{i.e.} a concurrently high $\sigmanorm$ and $\fpg$) which can form sufficiently massive seeds ($\geq\membstart$) inside the iceline contribute to the distribution inside $\sim2$ AU. The transition to higher  probabilities is not sharp as the iceline is itself a function of $\sigmanorm$ (fig. \ref{fig:rrockrice}).  Note that via the $\miso$ and $\tdisk$ criteria, the shape of the $\astart$ distribution  depends on the $\fpg$, $\sigmanorm$ and $\mwind$ distributions.

\subsection{Embryo start time $\tstart$}\label{subsection:embryostarttime} 
Depending upon initial conditions, the time needed to form just the initial seed embryo of 0.6 $\mearth$ is not short compared to the overall disk lifetime. We take this effect into account by introducing a time delay $\tstart$ between the beginning of the disk evolution and the time at which put the embryo into the disk. This time delay is calculated in a deterministic way as a function of $\fpg$, $\sigmanorm$ and $\astart$. It is therefore not an independent Monte Carlo variable. Similarly to $\astart$, the time delay cannot be constrained by observation. We therefore again rely on theoretical arguments. 

During the early stages of planetary accretion, the process of embryo growth proceeds mainly by accumulation of background planetesimals and can be  well described in the two-body approximation (Kokubo \& Ida \cite{kokuboida2002}). The accretion rate of an embryo growing from background planetesimals is thus  (e.g. Lissauer \& Stewart \cite{lissauerstewart1993})
\beq\label{eq:dmdtembryotstart}
\frac{d \memb}{dt}\simeq \pi R_{\rm emb}^{2} \Omega \sigmad \left( 1+\frac{v_{\rm esc}^2}{v_{\rm disp}^2} \right).
\eeq
Here, $R_{\rm emb}$ is the radius of the embryo,  $v_{\rm esc}$ the escape velocity from it and $v_{\rm disp}$ the typical random velocity of the planetesimals.

As the embryo grows, the solid surface density of planetesimals $\sigmad$ must decrease correspondingly  (Thommes et al. \cite{thommesetal2003}):
\beq
\frac{d \sigmad}{dt} = - \frac{(3 \mstar)^{1/3}}{6\pi \astart^2 B_L \memb^{1/3}}\frac{d \memb}{dt}\label{eq:tstartsigmadecrease}
\eeq
Looking at the dependence of the embryo growth time on semimajor axis, it is found (Steward \& Ida \cite{stewartida2000}; Weidenschilling \& Davis \cite{weidenschillingdavis2001}) that $\tstart$ increases significantly with distance to the star. The reason is that both the surface density of planetesimals and the frequency of collisions decrease with distance,  as the pace on which the later occur scales with  $\Omega$ (Kokubo \& Ida \cite{kokuboida2002}).  This increase of the growth timescale with distance is well reproduced in numerical simulations where a ``wave''  of embryo growth is seen to propagate from the inner parts to the outer  parts of the disk (Weidenschilling \& Davis \cite{weidenschillingdavis2001}; Thommes et al. \cite{thommesetal2003}).

Thus, building up a body of fixed mass, in our case the initial seed, takes longer for larger $\astart$,  a trend that holds everywhere but at the iceline where $\sigmad$  increases suddenly. To derive the corresponding $\tstart$, we integrate numerically equations \ref{eq:dmdtembryotstart} and \ref{eq:tstartsigmadecrease} until the mass has reached  $\membstart=0.6$ $\mearth$.  As in the formation model, we use the procedure of Pollack et al. (\cite{pollacketal1996}) to calculate the planetesimal random velocity $v_{\rm disp}$  and assume a constant planetesimal size of 100 km. As initial values at $t=0$, we use a mass of $0.66\mpla^{3/5}\miso^{2/5}$ (Chambers \cite{chambers2006}).

Compared to the formation model which calculates the evolution of the protoplanet starting with $\membstart=0.6$ $\mearth$, we neglect several things in the calculation of  $\tstart$: First, migration is neglected.  Tests have shown that with the type I migration efficiency factor $\f1$ we use for the nominal case,  migration of bodies  this small is indeed negligible. Second, we use just the simple particle in a box law for the gravitational focussing from eq. \ref{eq:dmdtembryotstart} instead of the full three body results  form Greenzweig \& Lissauer (\cite{GL92}), and third, we neglect the increase of the capture radius $\rcapt$ due to the gaseous envelope. The latter assumption is appropriate given the small seed masses and the assumed planetesimal size as shown by our own tests but also by Kornet \& Wolf (\cite{kornetwolf2006}) or Fortier et al. (\cite{fortieretal2007}).     

\begin{figure*}
     \begin{minipage}{0.49\textwidth}
      \centering
       \includegraphics[width=\textwidth]{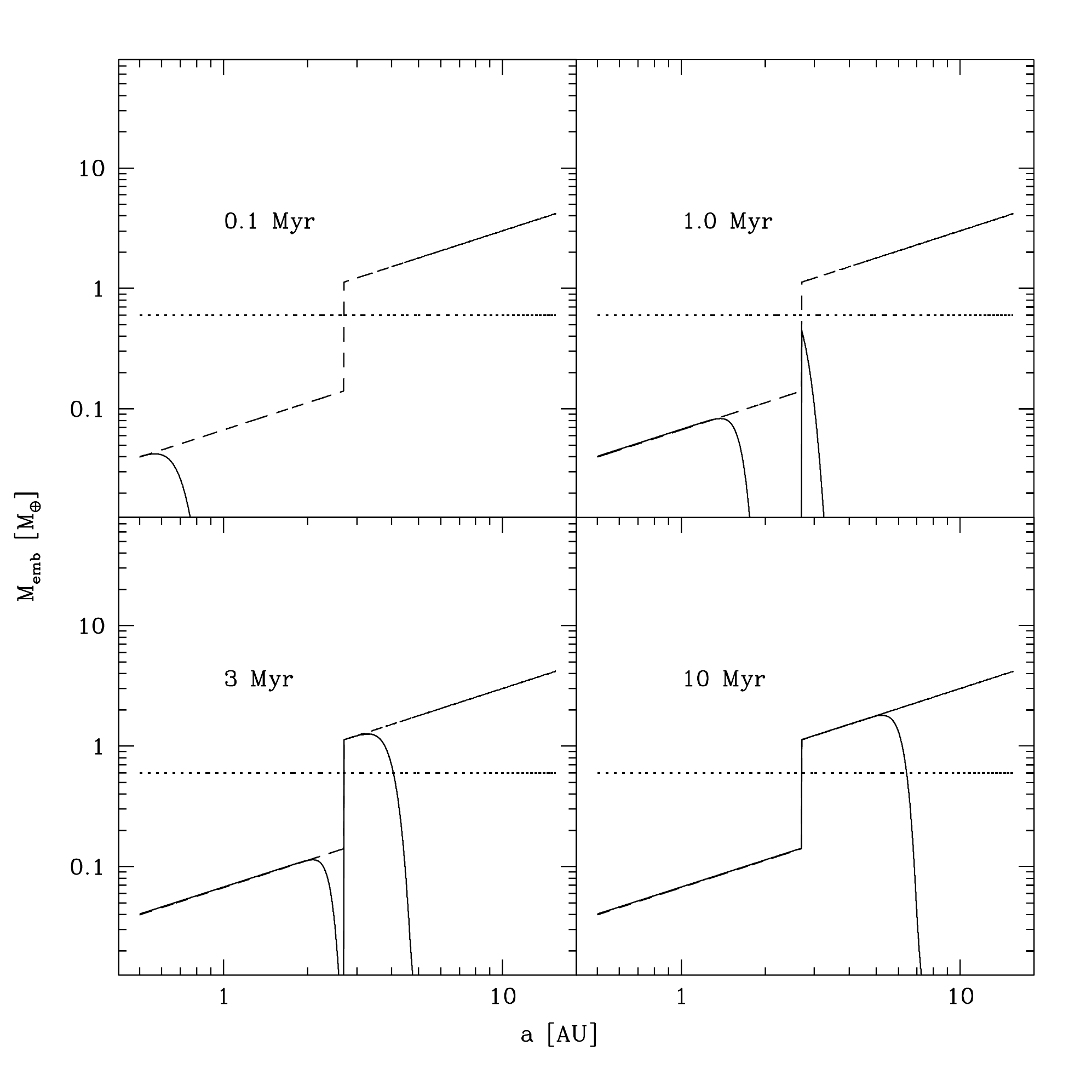}
     \end{minipage}\hfill
     \begin{minipage}{0.49\textwidth}
      \centering
       \includegraphics[width=\textwidth]{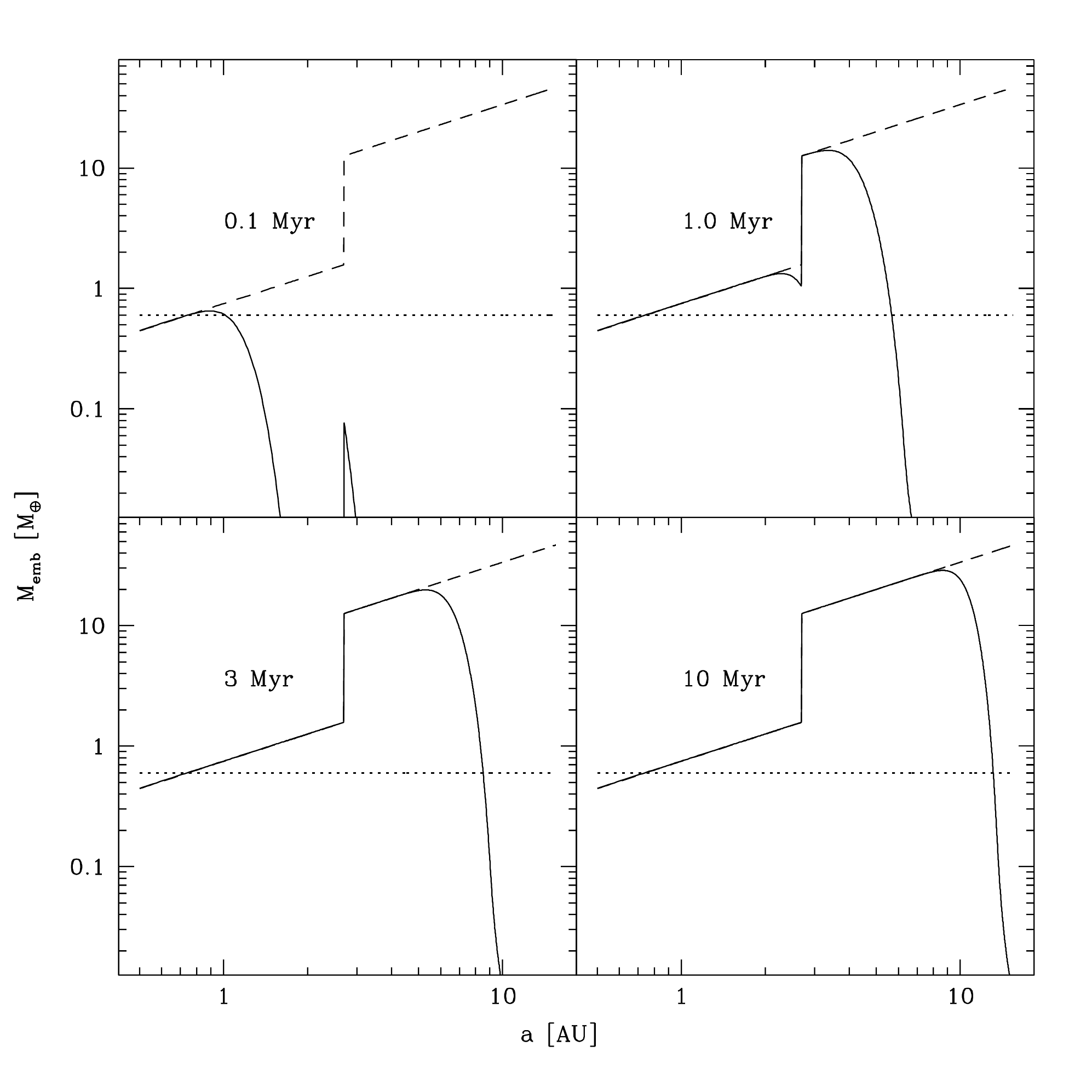}
     \end{minipage}
     \caption{Snapshots of the embryo mass (solid line) as a function
     of semimajor axis  at four moments in time for two different solid
     surface densities.  The dashed line is the isolation mass. The
     dotted line is $\membstart=0.6$ $\mearth$.  The initial solid
     surface density at 1 AU is 7 g/cm$^2$  (left panel) and 35
     g/cm$^2$ (right panel).  It should be kept in mind that this kind
     of calculation is just  needed to generate the start time
     $\tstart$ when the embryo is put into the formation model. The
     real evolution of the solid core for $M>0.6$ $\mearth$ is in
     general much more complex than plotted here.  In this figure, we have continued the
     calculations up to the isolation mass just to allow comparisons
     with other models.}
     \label{fig:atstart4snapshots}
\end{figure*}

Fig. \ref{fig:atstart4snapshots} illustrates the result of the numerical integrations to obtain $\tstart$. For a solid surface density at 1 AU of 7 g/cm$^2$ ($\sim$ MMSN) and  for  35 g/cm$^2$ ($\sim$ 5 x MMSN) four  snapshots in time of the embryo mass as a function of semimajor axis  are plotted. 

To allow comparisons with other models (Thommes et al. \cite{thommesetal2003};  Ida \& Lin \cite{idalin2004a}; Chambers \cite{chambers2006}),  we have  continued the integration up to $\miso$ for this plot instead of stopping at $\membstart$ as we  do when generating the initial conditions.  The dotted line in  fig. \ref{fig:atstart4snapshots} is $\miso$  and shows that for the MMSN, the earliest possible time to start is somewhat more than 1 Myr, just outside the iceline. Then, the domain of possible  start positions grows only slowly to larger semimajor axes.  After 10 Myr, when the gas disks will have disappeared (cf. fig. \ref{fig:remainingdiskfrac}), the largest possible start position  is still only 6-7 AU. In contrast, in the  5 x MMSN case, $\miso$ is larger than $\membstart$ inside the iceline too,  and the earliest embryo can start at around 0.8 AU slightly before 0.1 Myr.  At later times, embryos can start from all semimajor axes.  

Compared to the results of Thommes et al. (\cite{thommesetal2003}), our calculations show a faster growth of the cores, especially at large distances, which is due to the different way of computing planetesimal eccentricities and inclinations in their model, as illustrated by Fortier et al. (\cite{fortieretal2007}). Compared to Ida \& Lin (\cite{idalin2004a}), the results are quite similar, even if core growth proceeds at large orbital distances somewhat faster in our model. Compared to Chambers (\cite{chambers2006}) one finds that core growth in our model is faster than in his simple equilibrium model, but slower than in his complete model that is considerably more complex, including e.g. planetesimal fragmentation.

As mentioned in \S \ref{subsect:embryostarposition}, we only start embryos in that part of the disk where  $\miso\geq\membstart$ and $\tdisk\geq\tstart$.  The latter condition gives an outer bound for possible starting positions. The reasoning behind it is that if one of the numerous planetary seeds \textit{can} form while the disk is still present, it would indeed have done so, and that it is a candidate to eventually become a giant planet observable today. In other parts of the disk,  seed embryos also form, but they remain very small during the whole presence of the gaseous disk. Thus, we aim at minimizing the negative side effects of having only one seed per disk on the population of giant planets, but at the same time make our populations incomplete at small masses (cf. \S \ref{subsect:limitationsofthemodel}). 

For a significant fraction ($\sim 28$\%) of the sets of initial conditions we draw, one or both of the two aforementioned conditions cannot be fulfilled anywhere in the disk, namely when $\fpg$ and/or  $\sigmanorm$ come from the low tail of their distributions, while $\mwind$ is high. In such cases, no calculations were made, but we keep the record of the corresponding initial conditions where the formation of sizable planets is not possible and correct for them when calculating for example overall detection probabilities (paper II).  

\section{Results}\label{sect:results}
Once all Monte Carlo variables have been drawn, the next step consists in computing the formation of the planet corresponding to these initial conditions. This process can be illustrated by means of formation tracks in the mass-distance plane. Except where otherwise stated, all results are obtained for a population with  $\alpha=$0.007 and $\f1=0.001$. The reason for this choice is that the resulting sub-population of observable synthetic planets reasonably well reproduces the observed population (paper II).

\subsection{Planetary formation tracks}\label{subsect:exampleevopathes}
\begin{figure*}
      \centering
      \includegraphics[width=16cm]{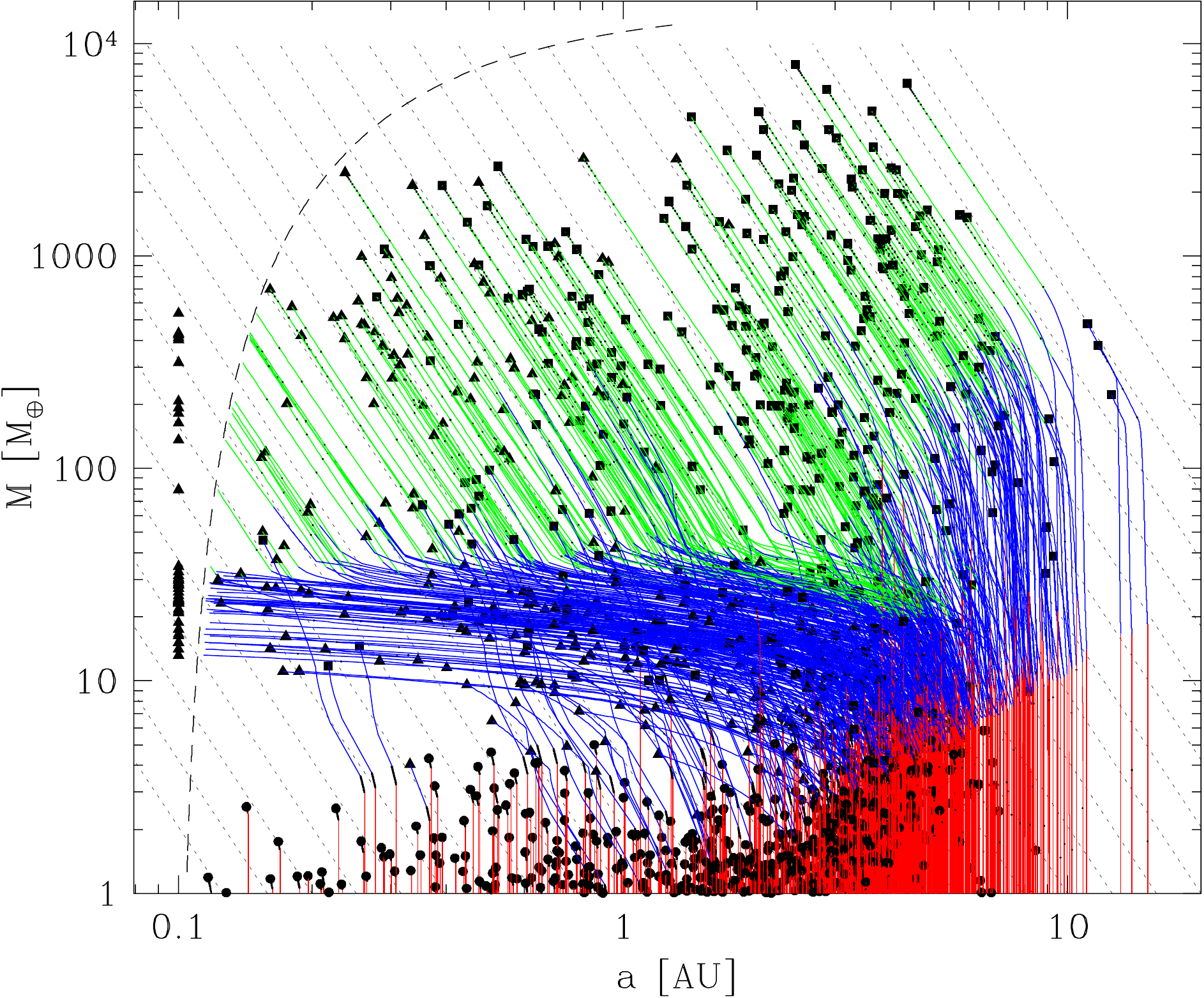}
      \caption{Planetary formation tracks in the mass-distance plane. The large black symbols show the final position of a planet. The shape of the symbols is explained in the text. Planets reaching the feeding limit at $\atouch$ (indicated by the long dashed line) have arbitrarily been set to 0.1 AU. The short dashed lines have a slope of $-\pi$ (discussion in \S \ref{subsubsect:mainclump}). Each track is color-coded according to the migration mode, and small black dots are plotted on the tracks all 0.2 Myr to indicate the temporal evolution of a planet.} 
      \label{fig:aMtracks}
\end{figure*}

Figure \ref{fig:aMtracks} shows formation tracks of  about 1500 randomly chosen synthetic planets. The tracks lead from the initial position at $a(t=0)=\astart$  and the fixed  $M(t=0)=\membstart$ to the final position marked by a large black symbol when planet growth and migration stops.  The color of the track indicates the migration mode:  Red for type I migration, blue for ordinary (disk dominated) type II migration and green for the braking phase. In this phase, planet dominated type II migration occurs (eq. \ref{eq:migrationrateduringbraking}) and the planetary gas accretion rate is given by the rate at which the disk can supply gas (eq. \ref{eq:dmdtdisklimited}). 

Even if the tracks show a great diversity, one can distinguish groups of planets with similar tracks. These groups are due to different formation stages that planets might undergo. In the next sections, we study representative tracks of four such groups.

\subsubsection{Tracks of ``Failed cores''}\label{subsubsect:failedcores}
During the first stage of formation at low masses type I migration (red) occurs. Since for this example population type I migration is very slow ($\f1=0.001$), the tracks are almost vertical. Planets that have migrated in type I only are represented by filled circles in fig. \ref{fig:aMtracks}.

For most embryos, this first stage is also the final one.  Their evolution stops at small masses because most initial conditions do not allow the formation of more massive planets during the lifetime of the disk. Therefore, most seeds (50-75 \%, see paper II) contribute to building up a large population of ``failed cores'' with $M \sim 1-10$ $ \mearth$ which, for a point of view of giant planet formation, failed to accrete a significant amount of gas. Also the population synthesis calculations of Ida \& Lin (\cite{idalin2004a,idalin2008}) contain a large sub-population of low mass planets. This is also compatible with the non-detection of giant planets around 90 to 95\% of nearby solar like stars. 

\begin{figure*}
   \begin{minipage}{0.49\textwidth}
      \centering
       \includegraphics[width=\textwidth,height=9.2cm]{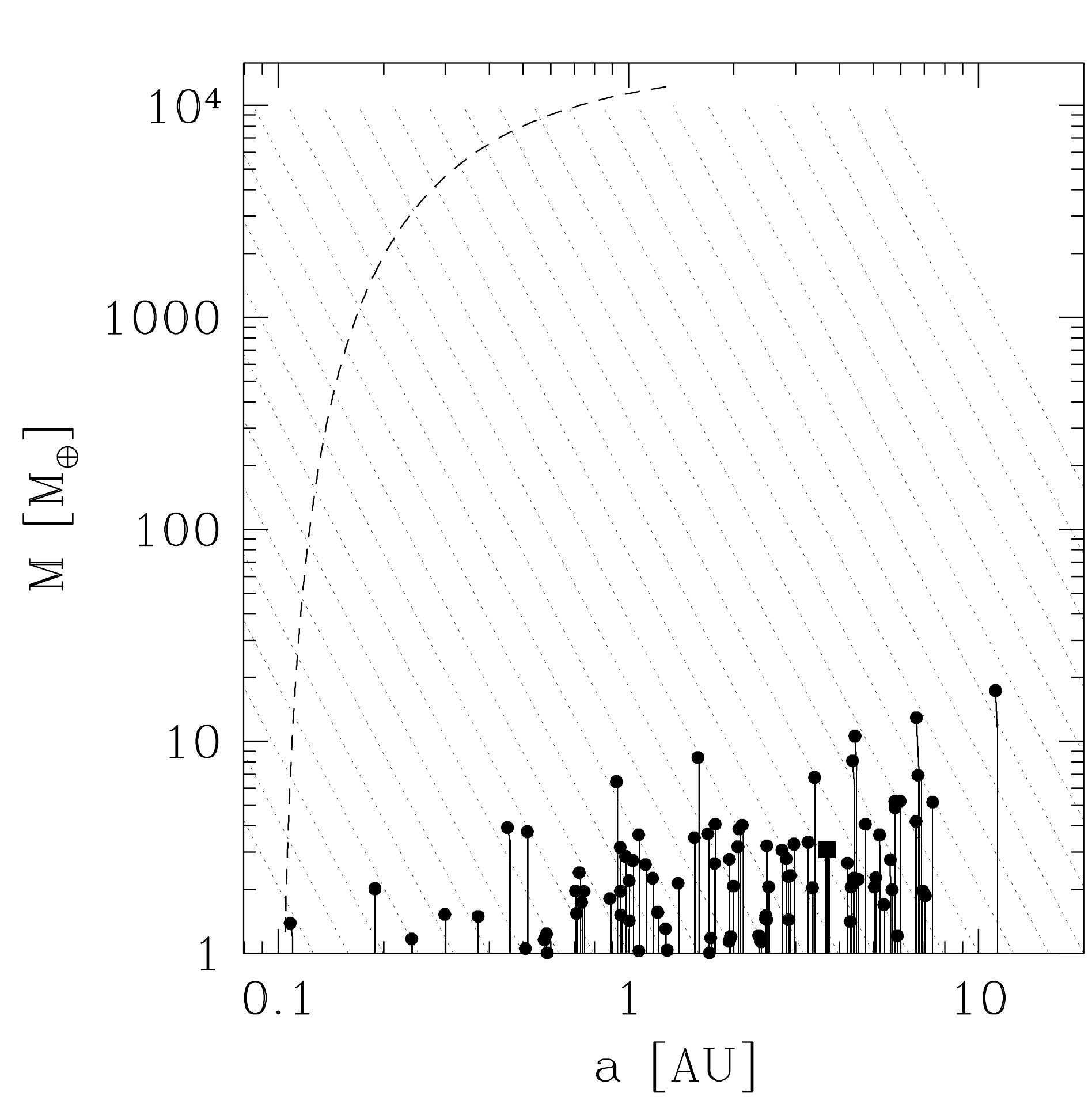}
    \end{minipage}\hfill
     \begin{minipage}{0.49\textwidth}
      \centering
       \includegraphics[width=\textwidth,angle=270]{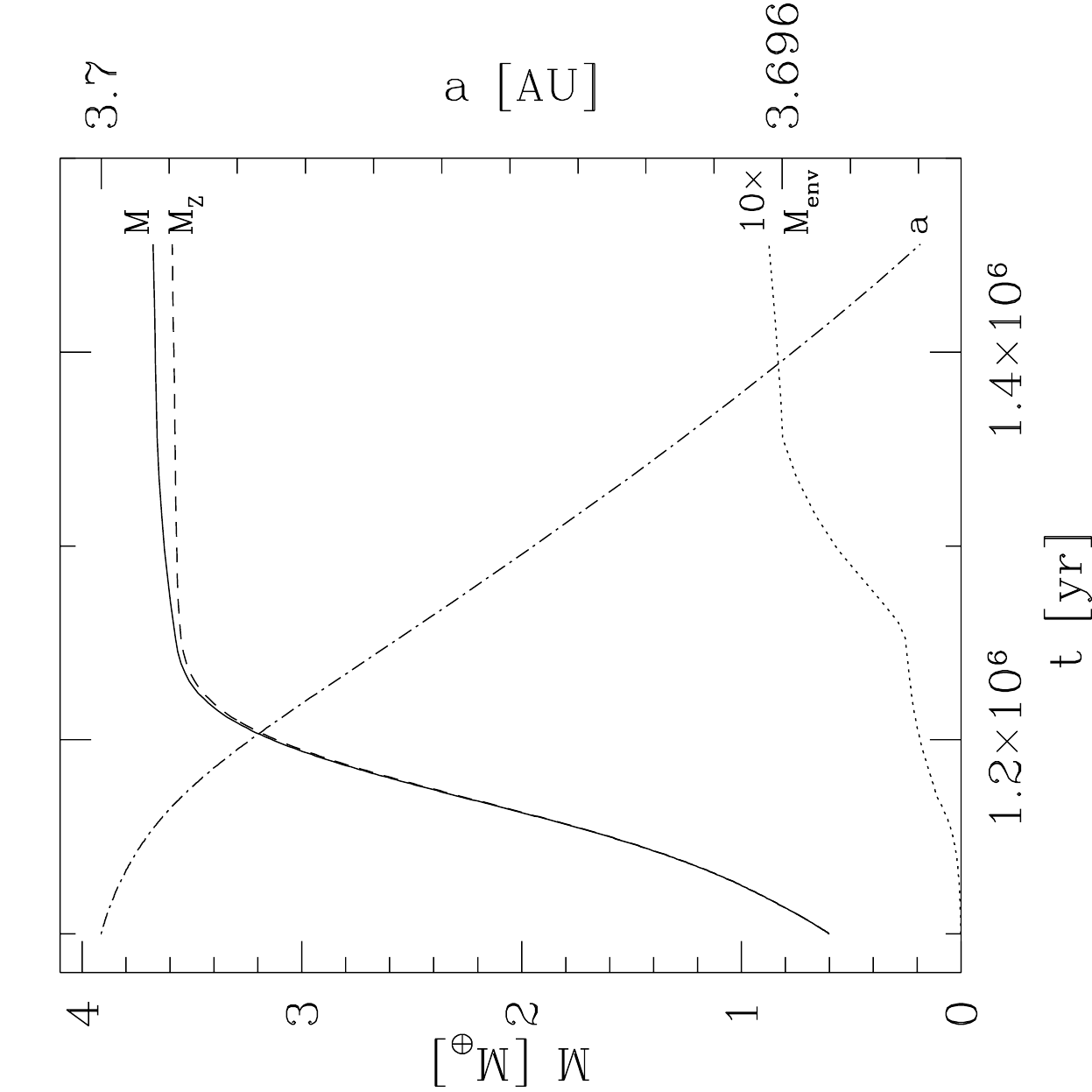}
     \end{minipage}
     \caption{Planetary formation tracks in the mass-distance plane for ``failed cores'' (left panel). The black points represent the final position of the planets. The thick line starting at 3.7 AU is the track of one prototypical example, for which the right panels shows the temporal evolution.  Its final position is represented by a large square.  In the right panel, the total mass $M$ (solid line), the mass of accreted solids $\mheavy$ (dashed line) and the mass of the envelope $\menv$ multiplied by a factor 10 for better visibility (dotted line) are plotted (scale on the left) as a function of time $t$.  The temporal evolution of the planet's semimajor axis $a$ is also plotted (dash-dotted line, scale on the right). }
\label{fig:prototypsFC}
\end{figure*}

In fig. \ref{fig:prototypsFC}, left panel, exemplary formation tracks for a number of such planets are plotted. As expected from eq. \ref{eq:isomassspec} for $\miso$, ``failed cores'' can reach larger masses at larger distances. The right panel of fig. \ref{fig:prototypsFC} shows the temporal evolution of the mass and semimajor axis of one typical ``failed core''. This seed starts at $\astart=3.7$ AU in a disk with $\fpg=0.028$ ([Fe/H]=-0.15) and $\sigmanorm=165$ g/cm$^{2}$. This initial position is situated not far outside the iceline.  For such a solid surface density, forming the initial seed takes a significant amount of time (cf. fig. \ref{fig:atstart4snapshots}), namely about 1.1 Myr. 

 As is shown in the right panel of fig. \ref{fig:prototypsFC}, the core then quickly accretes all  planetesimals in its reach. Gas accretion is of negligible importance. At about 1.2 Myr, the mass of the core approaches the local isolation mass\footnote{From eq. \ref{eq:isomassgeneral} one would calculate a $\miso$ of about $2.8 \mearth$, using $B_{L}=4$. However, as we do not reduce the initial solid surface density by the amount of material already in the initial seed, a value larger for the mass by about $3/2\times\membstart$, is obtained.}.  For the remaining 0.2 Myr of evolution, the core grows only very slowly. One notes that the envelope now becomes somewhat more massive, due to the reduced luminosity of the core. Clearly, the evolution of this planet simply corresponds to the two first phases described by Pollack et al. (\cite{pollacketal1996}), with the difference that further evolution is inhibited by the dispersion of the protoplanetary nebula after 1.45 Myr. At this time, we are left with a ``failed core'', consisting of about $3.6$ $\mearth$ of heavy elements, and $\sim0.1$ $\mearth$ of gas. The extend over which migration occurred is tiny because of $\f1=0.001$, roughly 0.004 AU, much less than the extent of the planet's Hills radius.  The fact that further growth is simply inhibited by the disappearance of the gaseous disk is characteristic for this type of planet.

The vast sub-population of ``failed cores'' is not identical to the \textit{final} terrestrial planet population, expected to be located in a similar $a-M$ region. Rather, they represent an earlier moment in evolution. ``Failed cores'' are formed from one large embryo accreting small field planetesimals while the gas disk is still present. Terrestrial planets on the other hand get their final properties from giant impacts between bodies of a similar size (several ``failed cores'') on much longer timescales, a phase missing in our model. We expect that after disk dispersal, all the ``failed cores'' of one disk would start to interact gravitationally, leading to scattering, ejections and collisions, until the remaining planets have settled into stable orbits (e.g. Ford \& Chiang \cite{fordchiang2007}; Thommes et al. \cite{thommesetal2008}).

\subsubsection{Tracks of ``horizontal branch'' planets} \label{subsubsect:horizontalbranch}
In some other cases the core grows so large (and does so sufficiently quickly) that the planet can open a gap in the gas disk long before the latter disappears.  At this point, the migration mode changes to disk dominated type II (blue lines in fig. \ref{fig:aMtracks}), which is the second phase.

After a short transitional phase, planets starting inside about 4-6 AU begin then to move in disk dominated type II migration along nearly, but not completely horizontal tracks at $M\sim 7-30$ $\mearth$. These nearly horizontal tracks are clearly seen in fig. \ref{fig:aMtracks} forming a ``horizontal branch'' of planets. We identify the planets having had such phase during their formation a posteriori by the condition that while in type II migration, one finds $\frac{d\log M}{d\log a}<0.1$. Physically this means that  migration occurs on a significantly shorter timescale than accretion. Planets having passed through the ``horizontal branch'' have their final position marked by triangles in fig.  \ref{fig:aMtracks}. 

\begin{figure*}
   \begin{minipage}{0.49\textwidth}
      \centering
       \includegraphics[width=\textwidth,height=9.15cm]{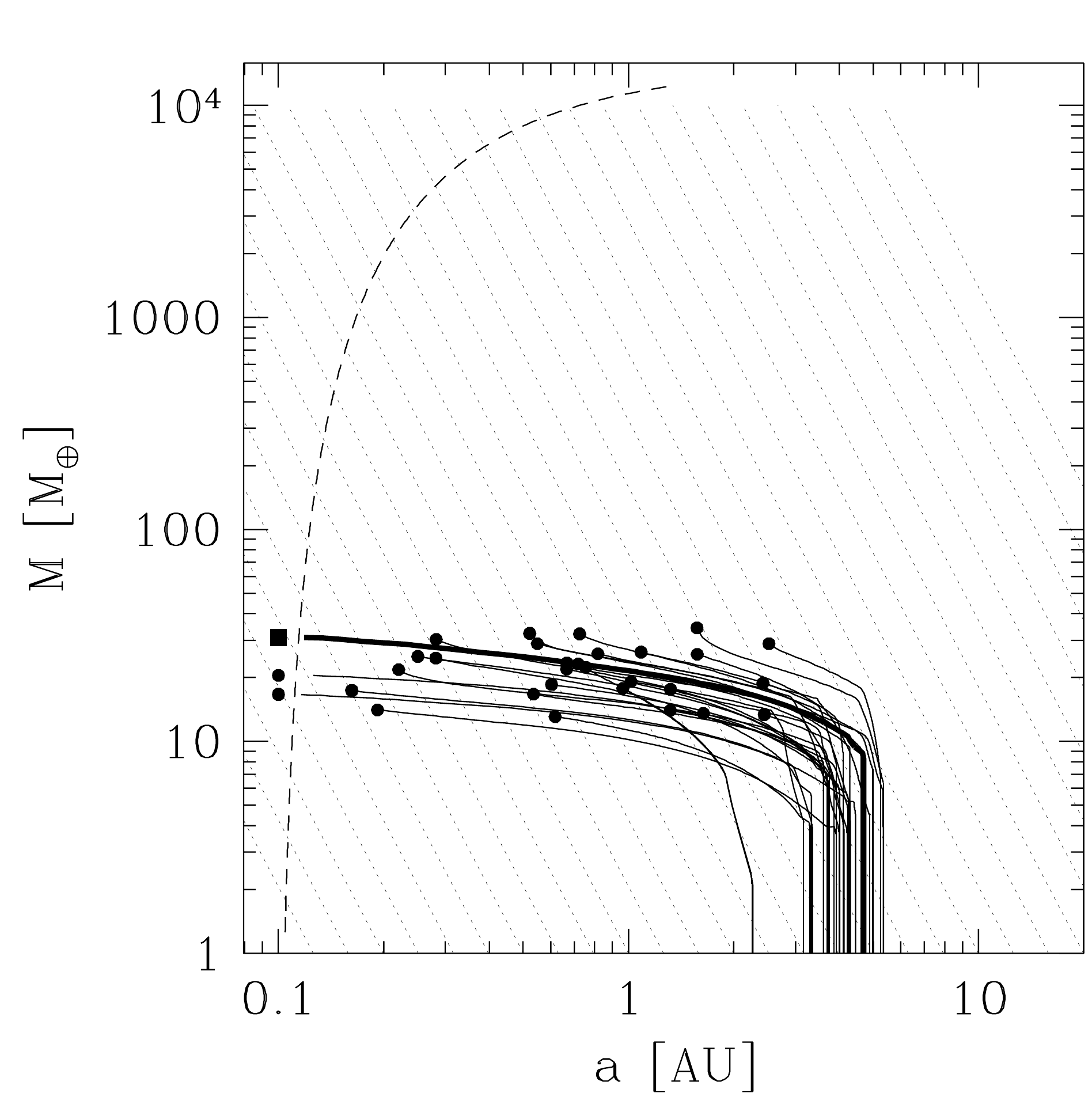}
    \end{minipage}\hfill
     \begin{minipage}{0.49\textwidth}
      \centering
       \includegraphics[width=\textwidth]{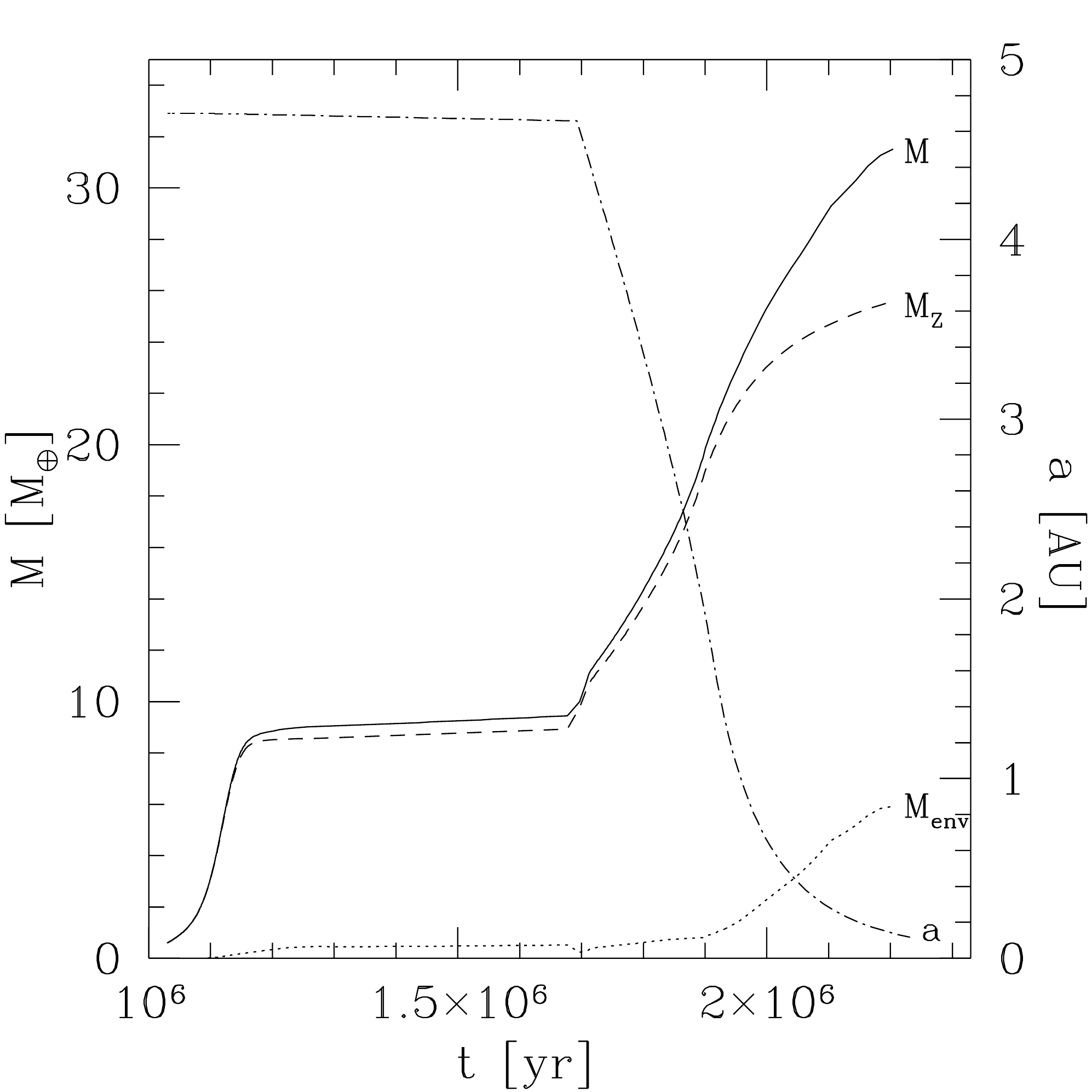}
     \end{minipage}
     \caption{As fig. \ref{fig:prototypsFC}, but for planets of the ``horizontal branch''. In this case, the prototypical track (thick line, and large square at the final position) starts at $\astart=4.7$ AU. It ends as a ``Hot Neptune'' planet in the feeding limit at 0.1 AU.  Its temporal evolution is shown in the right panel.}
\label{fig:prototypsHB}
\end{figure*}

Figure \ref{fig:prototypsHB}, left panel, shows some exemplary formation tracks of planets that stay in the ``horizontal branch'' until the gas disk has disappeared (so that they end up at intermediate distances),  or until they reach the feeding limit  (so that they have a final position $\lesssim0.1$ AU). 

The prototypical example of the right panel of fig. \ref{fig:prototypsHB} is formed in a disk whose mass is similar to the mean values adopted in this work ($\sigmanorm=270$ g/cm$^2$). The dust to gas ratio is with $\fpg=0.03$ somewhat smaller than the mean value. This leads to a formation time of the initial seed of just about 1 Myr. For this disk, the iceline is located at 4.2 AU, therefore the planet accretes icy planetesimals at the beginning of its formation. During the first 1.6 Myr, the formation is similar to the one described in classical core accretion papers, namely a rapid core formation (up to about 8 $\mearth$, the isolation mass, in 0.15 Myr) followed by a phase of low mass growth, similar to phase 2 of Pollack et al. (\cite{pollacketal1996}). Just before $t=1.7$ Myr however, migration switches to type II, due to the concurrent  growth of the planet and the decrease of the disk scale height with time, so that also a planet of a relatively low mass can open a gap in the disk. 

In the population presented here, we have reduced type I migration by a large factor. Therefore changing from the (strongly reduced) type I to (normal) type II results in a net  increase of the migration rate and the planet moves into regions of the disks that have not yet been depleted in planetesimals  (Alibert et al. \cite{alibertetal2004}). This significantly increases the core growth and and hence its luminosity. The latter translates in a slight decrease of the envelope mass at 1.7 Myr.

Shortly after switching migration type, the planet crosses the iceline, which reduces the solid accretion rate (see the little kink on the mass lines just after 1.7 Myr).  During the remaining accretion inside the iceline, another $\sim17$ $\mearth$ of rocky planetesimals are collected. Comparing this to the total amount of heavy elements initially inside the iceline  ($\sim 20$ $\mearth$) shows that the planet is quite efficient in emptying the planetesimal disk. As a consequence, the final core mass is approximately the sum of the mass collected while passing through the ``horizontal branch'', plus the mass of the icy planetesimals accreted during the initial in-situ growth phase. 

At roughly 1.9 Myr, the local disk mass becomes comparable to the planet's mass so that the migration rate starts to slow down (fig. \ref{fig:sigmaaalocaldiskmass}) and the planets starts to accrete gas. However, migration continues (albeit at a reduced rate) and the planet enters at $t=2.2$ Myr into the feeding limit at roughly 0.1 AU, where we stop the calculations. 

Contrary to the two remaining types of representative tracks described below, the gas accretion rate of ``horizontal branch'' planets is governed during their entire formation by the planet \textit{i.e.} the maximum accretion rate limited by the disk is never reached. 
Since the gas accretion rate is moderate (due to the fact that the core luminosity is most of the time quite large), the planet at the end of its formation is made of a large core ($\sim26$ $\mearth$) and a significant, but still much smaller envelope ($\sim6$ $\mearth$). We find that the sub-population of close-in, low mass planets ($M\lesssim30-40$ $\mearth$) is characterized by a ratio $\menv/\mcore$ that varies between $\sim$0.02 (at $M\sim10\mearth$) to $\sim$ 0.3  (at $M\sim40$ $\mearth$), \textit{i.e.} these planets have a structure roughly comparable to Neptune. This could be different for close-in Neptune mass planets formed through giant impacts between initially smaller bodies (Ida \& Lin \cite{idalin2008}), especially as the impacts could remove their anyway tenuous gaseous envelopes. Transiting Neptune mass planets around solar like stars can serve as a diagnostic to distinguish these different formation channels. The recently detected transiting ``Hot Neptune'' HAT-P-11b (Bakos et al. \cite{bakosetal2009}) has a radius compatible with a rock/ice core with a 10\% H/He envelope, whereas a pure rock/ice planet (as well as a miniature gaseous planet) are excluded (Bakos et al. \cite{bakosetal2009}). This suggest that this planet was formed in a similar way as described here. 

The ''horizontal branch'' is thus the ''conveyor belt'' by which Neptune-like planets are being transported close to the star.  These ``Hot Neptunes'' are a sub-population of planets that high precision radial velocity surveys (using e.g. HARPS, see Lovis et al. \cite{lovisetal2006}) now find in increasing numbers. Note that subsequent evolutionary effects, namely evaporation can sometimes significantly modify the structure of these bodies (Baraffe et al. \cite{baraffeetal2006}).  
  
\subsubsection{Tracks of ``main clump'' planets}\label{subsubsect:mainclump}
The fact that the tracks in the ``horizontal branch'' are not completely horizontal, \textit{i.e.} that growth in mass continues in this phase is important for the further evolution of the third group of planets, the planets of the ``main clump''. 

These are planets with final masses mostly between the mass of Saturn and three Jupiter masses, and semimajor axes mainly between $\sim0.3$ and 2 AU (see fig. \ref{fig:aM}).  For these planets, the core grows to a size that triggers runaway gas accretion while the planet is collecting solids as it passes through the ``horizontal branch''. Once runaway is triggered, the gas accretion rate increases very rapidly. With its rapidly growing mass, the planet soon exceeds the local disk mass and migration enters the planet dominated type II regime. The planet leaves the ``horizontal branch'' upwards in mass thereby starting its third phase of formation (plotted in green in fig.  \ref{fig:aMtracks}) .

\begin{figure*}
   \begin{minipage}{0.49\textwidth}
      \centering
       \includegraphics[width=\textwidth,height=9.3cm]{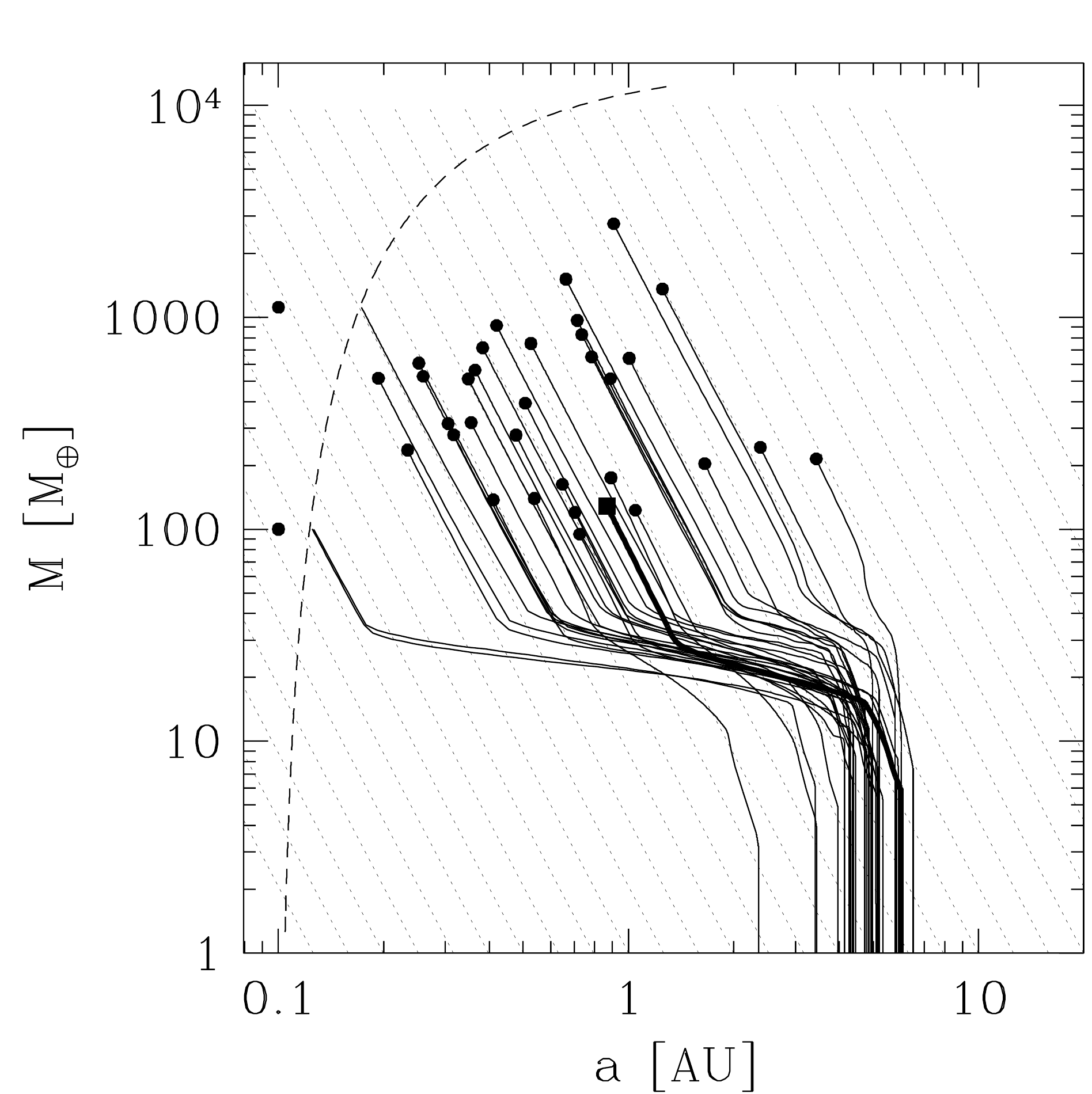}
    \end{minipage}\hfill
     \begin{minipage}{0.49\textwidth}
      \centering
       \includegraphics[width=\textwidth]{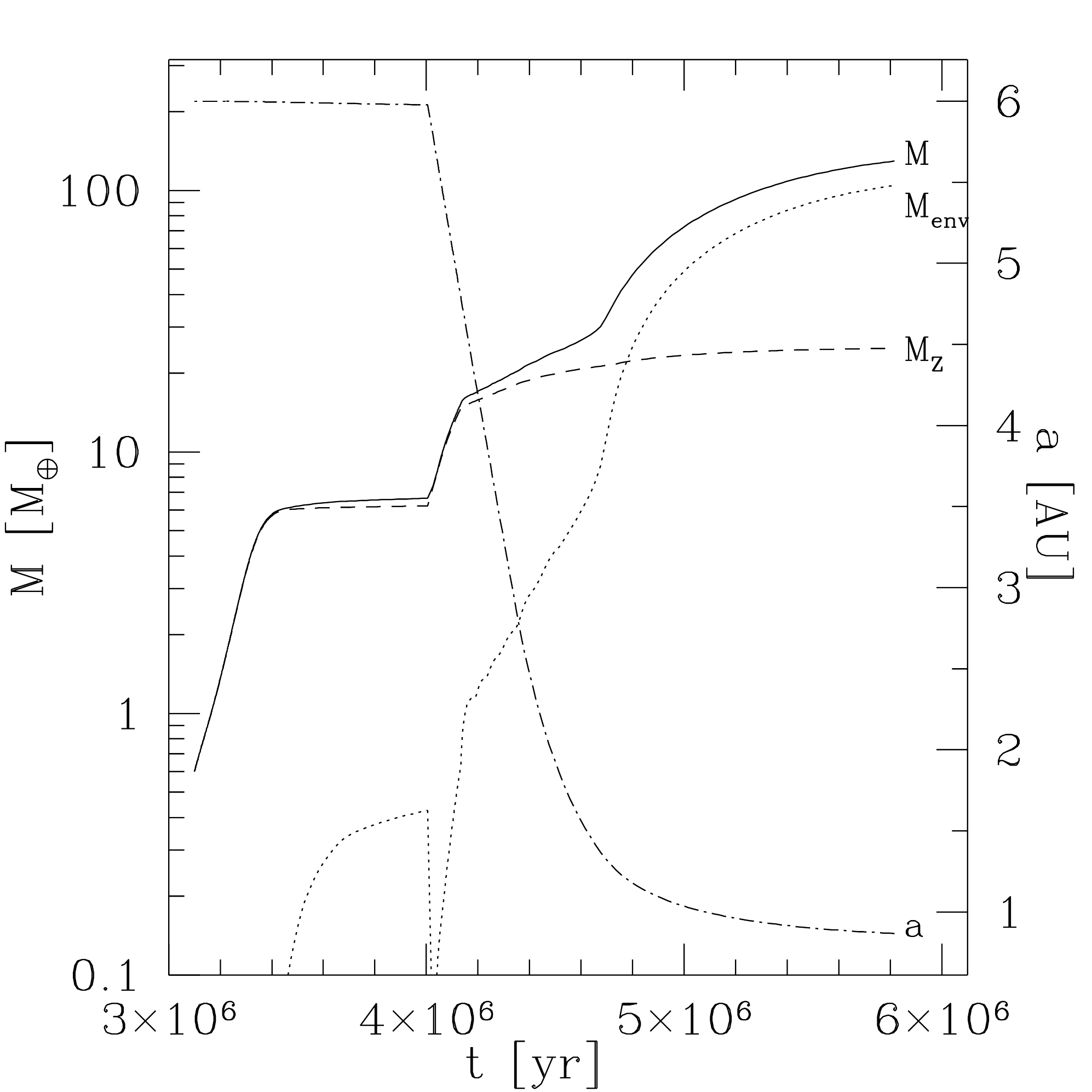}
     \end{minipage}
     \caption{As fig. \ref{fig:prototypsFC}, but for planets passing through the ``horizontal branch'' to become members of the ``main clump''. Here, the prototypical embryo (thick line, square at the final position, temporal evolution in the right panel) eventually leads to a Saturnian planet situated at 0.9 AU.}
\label{fig:prototypsMC}
\end{figure*}

During this final braking phase, the planets migrate at the reduced type II rate (eq. \ref{eq:migrationrateduringbraking}), and accrete gas at a rate given by the disk's evolution (eq. \ref{eq:dmdtdisklimited}). This has the interesting consequence that if we combine these equations, we find that planets move on formation tracks which are in the $\log(a)-\log(m)$ plane straight lines with a slope $d \log M/d \log a= -\pi$. This behavior is clearly visible in the formation tracks.  In fact, we have used the criterion $\vert\ d\log M/d\log a+\pi\vert<0.1$ to identify the braking phase a posteriori.  The evolution along these straight lines slows down in time as can be seen by the increasing number of small black ticks on the track near the final position of the planet (fig. \ref{fig:aMtracks}). This is a consequence of the concurrent decrease of the gas accretion rate due to disk evolution, and the slowing down of migration. At the end, the planet has accreted all the outer disk gas that has not been photo-evaporated. 

Examples of planets that undergo such a three staged evolution are plotted in figure \ref{fig:prototypsMC}. Two specific planets have reached the feeding limit near 0.1 AU, illustrating how Pegasi planets form, even though our model does not further treat the formation of ``Hot'' planets once they have migrated to $\atouch$.


The prototypical planet of the ``main clump'' in the right panel of this figure is formed in a disk with $\fpg$=0.02, \textit{i.e.} [Fe/H]=-0.3, and $\sigmanorm=280$ g/cm$^2$. The planet starts its formation beyond the iceline ($\astart=6$, $\aice=4.3$ AU), and the beginning of its formation (up to 4 Myr) is similar as for the ``horizontal branch'' planet before: the planets empties its feeding zone reaching an isolation mass of about 6 $\mearth$. Just before $t=4$ Myr, migration switches to type II. Shortly thereafter at 4.1 Myr the planet crosses the iceline (see the prominent kink on the solid line in fig. \ref{fig:prototypsMC}). Finally, the braking phase starts at about 4.4 Myr. Switching from type I to type II migration results in an increase of both the solid accretion rate and the core luminosity with a corresponding loss of envelope. The difference to the previous case originates from a longer living disk and the fact that the planet started at a larger distance from the central star. The solid mass available while migrating through the ``horizontal branch'' is therefore larger (it scales with $r^2 \sigmad \propto r^{0.5}$) and the core reaches a mass large enough (21 $\mearth$) to trigger a runaway accretion of gas around 4.6 Myr, significantly before the gas disk disappears (or the planet reaches the feeding limit as for the ``horizontal branch'' case). The gas accretion rate then increases rapidly, but soon (at 4.7 Myr) reaches the maximum rate allowed by the disk which is a rather moderate $1.6\times10^{-4}$ $\mearth$/yr. The disk limited accretion rate then decreases slowly with time as the disk evolves. From 4.7 Myr onwards the planet is in the braking phase with the characteristic $-\pi$ slope in the $a-M$ diagram until the disk disappears at 6 Myr. At this point, a Saturnian planet has formed at 0.9 AU with a total mass of about 130 $\mearth$ and a $M_{\rm Z}\approx 25 \mearth$. The prototypical planet also illustrated how the combination of effects that are per se straightforward (changing the migration regime, crossing the iceline, disk limited gas accretion rate) can lead to a complex formation history. 

It is interesting to note that the  ``main clump'' region of the $a-M$ diagram is somewhat overpopulated in the observed population as well. Examples are (among many others) \object{HD\,100777b} (Naef et al. \cite{naefetal2007}) or \object{HD\,142b} (Tinney et al. \cite{tinneyetal2002}).

\subsubsection{Tracks of ``outer group'' planets}\label{subsubsect:outergroup}
For starting positions larger than $\sim$ 4-7 AU, the formation tracks are different. In the case a high $\fpg$ and $\sigmanorm$ and a low $\mwind$ are drawn together, the core growth timescale is small compared to the disk depletion timescale even at these large distances. As the amount of solid material available is large, embryos can then grow to a supercritical mass almost in-situ (nearly vertical tracks up to several tens of $\mearth$ in  fig. \ref{fig:aMtracks}), without the need of collecting solid material by migration. Significant migration can nevertheless occur but only in the planet dominated type II mode and when the gas accretion rate is regulated by the disk. Such planets which do not have a ``horizontal branch'' phase  are represented in fig. \ref{fig:aMtracks} by filled squares. The final semimajor axes of the in-situ supercritical planets are outside $\sim$ 0.4-1 AU, but overlap with ``main clump'' planets. Sometimes ``outer group'' planets become extremely massive ``Super Jupiters'', with masses more than one order of magnitude larger than that of Jupiter.

\begin{figure*}
   \begin{minipage}{0.49\textwidth}
      \centering
       \includegraphics[width=\textwidth,height=8.9cm]{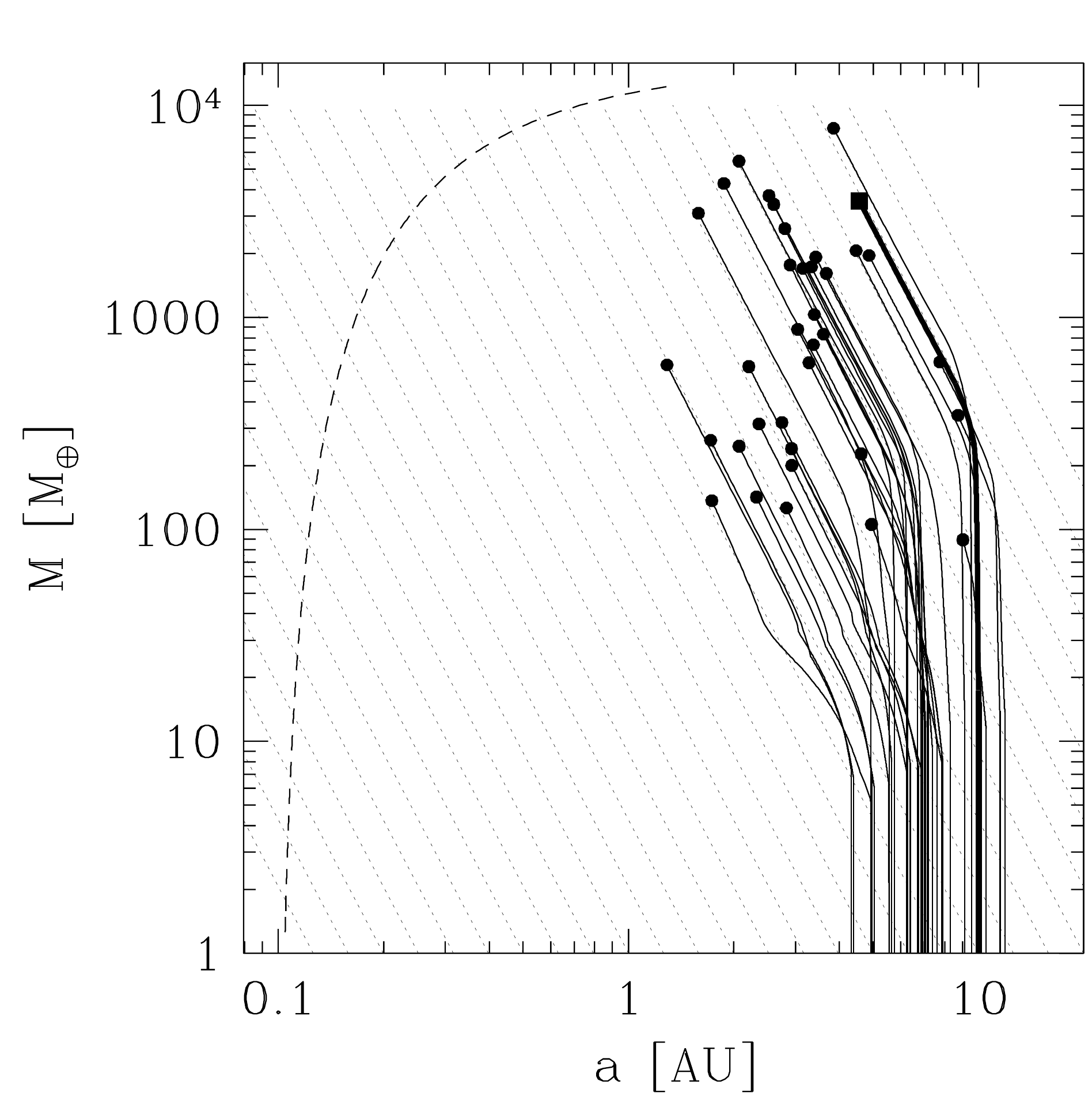}
    \end{minipage}\hfill
     \begin{minipage}{0.49\textwidth}
      \centering
       \includegraphics[width=\textwidth]{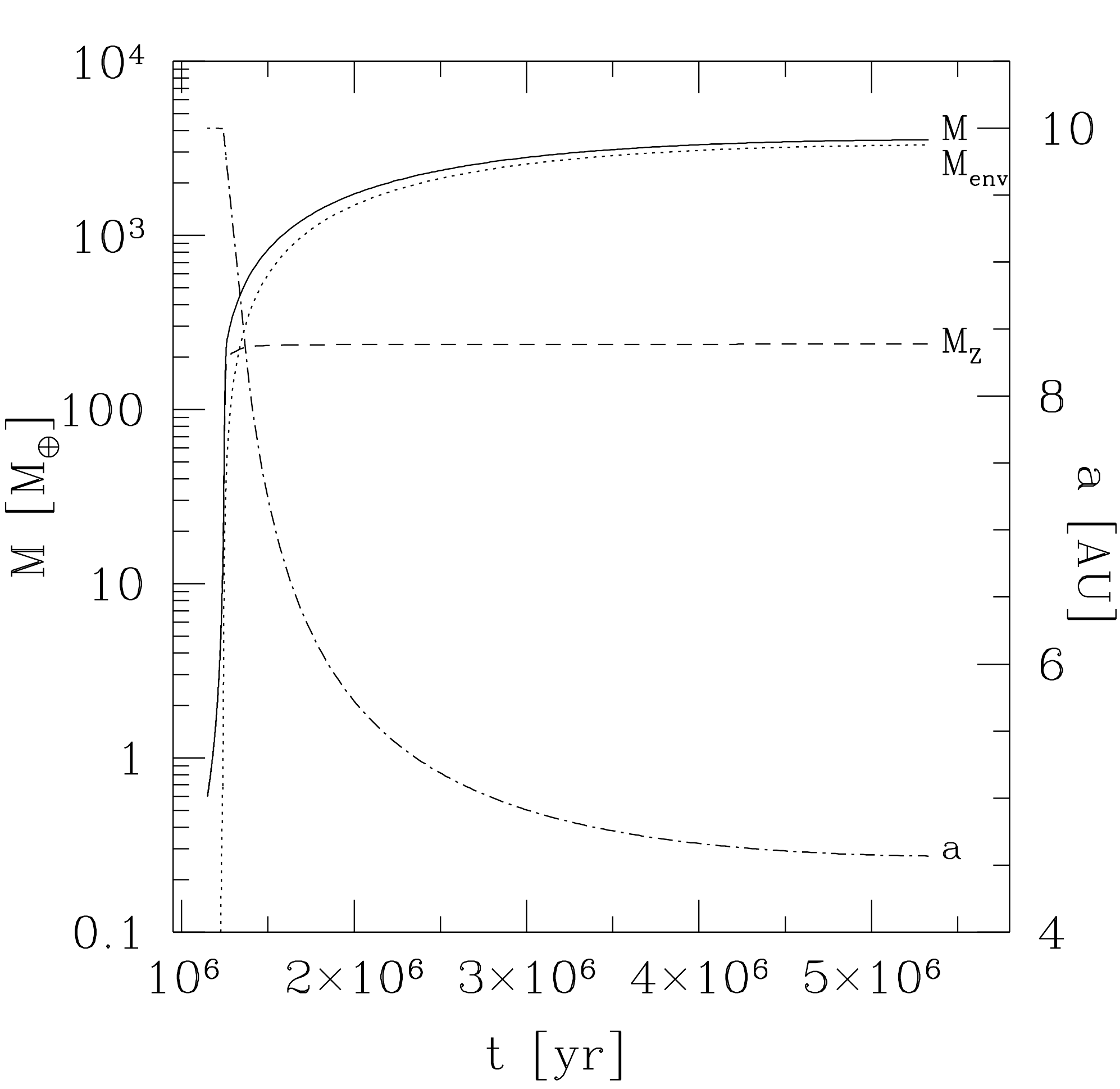}
     \end{minipage}
     \caption{As fig. \ref{fig:prototypsFC}, but for in-situ critical cores which end up in the ``outer group''. The prototypical embryo starts in a massive, metal rich, long lived disk which allows the formation of a very massive $\sim11\mj$ ``Super Jupiter'' ultimately located at 4.5 AU.}
\label{fig:prototypsOG}
\end{figure*}

Examples of  such tracks are plotted in figure \ref{fig:prototypsOG}. Comparing these tracks with the ones of the ``main clump'' shows that there is a continuous transition between the two types. The prototypical ``outer group'' planet in the right panel of fig. \ref{fig:prototypsOG} is formed in a massive, metal-rich ($\sigmanorm=500$ g/cm$^{2}$, [Fe/H]=0.3) and long lived disk. Due to the large $\astart$ (10 AU), the isolation mass at the initial location is then very large (around 150 $\mearth$). The seed starts at 1.1 Myr,  and rapidly switches to type II migration at 1.2 Myr. The critical mass (around 25 $\mearth$) is attained already at 1.24 Myr at a position very close to $\astart$, triggering a rapid accretion of gas. At 1.25 Myr, the gas accretion rate reaches the value limited by the disk of initially initially $2.5\times10^{-3}$ $\mearth$/yr.

Note that the final amount of heavy elements in this planet is quite uncertain. Indeed, ``outer group'' planets undergo a large part of their formation history in the disk limited gas accretion phase. As mentioned above, the internal structure of planets during this phase is no more calculated, thus their solid accretion rate (via $\rcapt$) is uncertain. As a consequence, the final $M_{\rm Z}$ found here (around $250$ $\mearth$) is unsure, and only a lower boundary ($\sim 25$ $\mearth$) is actually well determined. Internal structure models (Baraffe et al. \cite{baraffeetal2008}) do however require comparable amounts of heavy elements to reproduced the observed mass-radius relation of the transiting ``Hot Super Jupiter'' \object{HD\,147506b} (Bakos et al. \cite{bakosetal2007}). Thus, even if this planet is in terms of semimajor axis very different than the planet discussed here, it still illustrates that objects with very large amounts of solids in their interior seem to exist.  Figure \ref{fig:prototypsOG} shows that the forming planet is eventually strongly dominated by gas (about 10 $\mj$ are accreted). Therefore, the uncertainty on the final {\it total} mass (which is of primary interest in this study) is anyway low.

The figure also shows that $M_{\rm Z}$ (with all the uncertainties discussed above) reaches its final value between 8 and 9 AU already. The planet could in principle continue to accrete all the planetesimals down to its final location at 4.5 AU. However, the planet is so massive at this stage that nearly all planetesimals are actually ejected and not accreted (Ida \& Lin \cite{idalin2004a}). We however find that the large size of forming planets makes them less effective in ejecting planetesimals than old, compact objects of the same mass.

\subsection{Mass-semimajor axis diagram}\label{subsect:masssemimajoraxisdiagramm}
The planetary formation tracks illustrate how planets grow in mass and migrate to their final position. In this section, we discuss some aspects of the distribution of final masses and positions.

\subsubsection{Diversity of planets}\label{subsubsect:masssemimajoraxisdiagramm}
\begin{figure*}
      \centering
    \includegraphics[width=16cm]{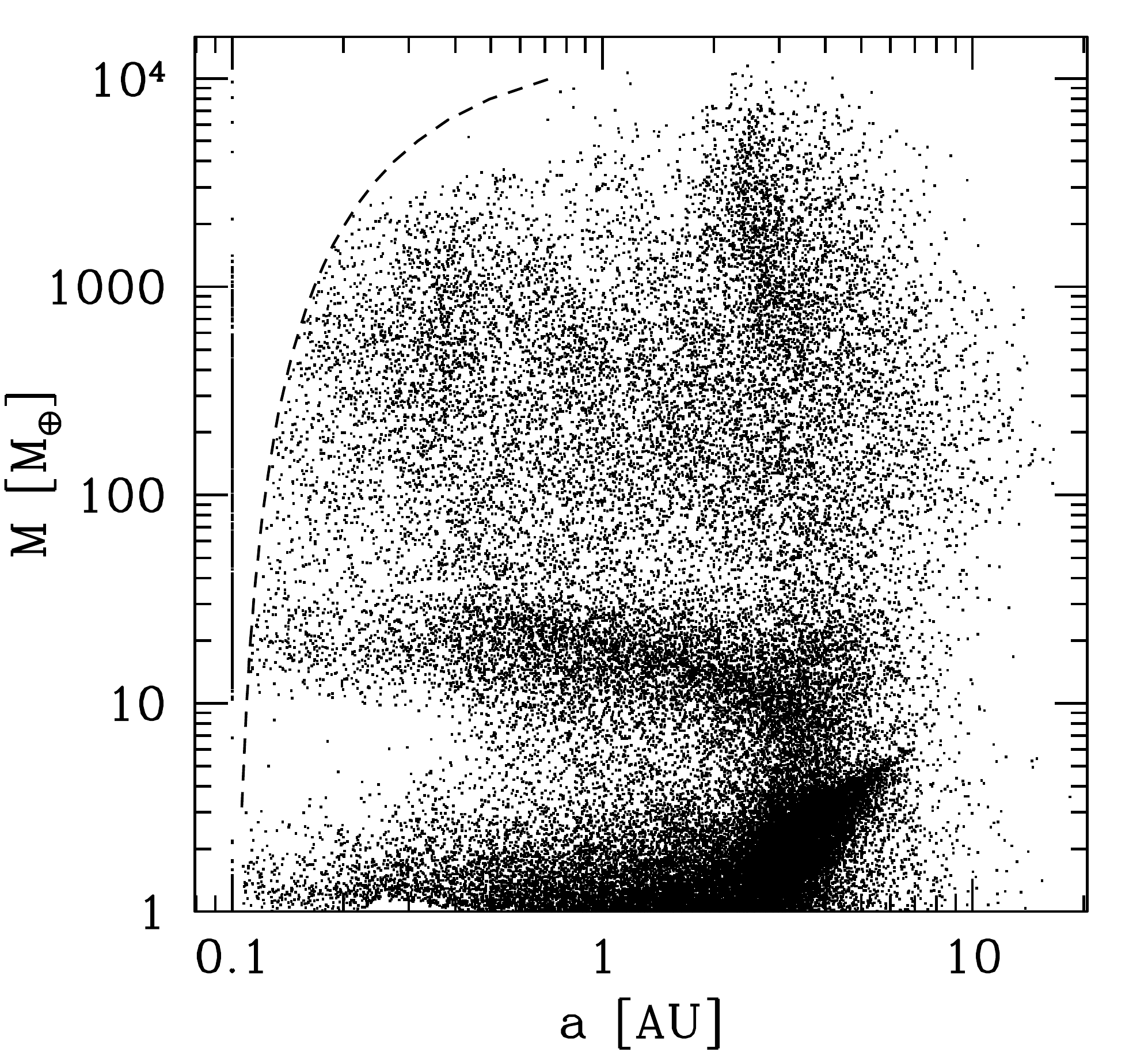}
      \caption{Final mass $M$ versus final distance $a$ of $\nsynt\approx50\,000$ synthetic planets of the nominal planetary population.  The feeding limit  at $\atouch$ is plotted as dashed line. Planets migrating into the feeding limit have been put to 0.1 AU. As $\atouch$ gets very large for $M\gtrsim20\mj$, also a few extremely massive planets are in the feeding limit which should however be regarded as a simulation artifact because our simplification of putting planets that reach the feeding limit to 0.1 AU ceases to be justified.} 
      \label{fig:aM}
\end{figure*}

Figure \ref{fig:aM} shows the mass and the semi-major axis of $\nsynt\approx50\,000$ synthetic planets. The first striking result is that core accretion allows for a very diverse planet population. The final mass of the planets varies between the smallest possible mass (0.6 $\mearth$) and a few extremely massive $\sim40$ $\mj$ planets. The final position varies between the innermost possible radius ($\approx0.1$ AU) to about 18 AU. We conclude that the observed diversity of extrasolar planets is within the core accretion paradigm a natural consequence of the observed diversity of the properties of the protoplanetary disks, which we have varied within the observed range. It is therefore obvious that a better understanding of the properties of the protoplanetary disks, especially the innermost region, has important implication for this planet formation theory.  In our population synthesis, a number of parameters are kept fixed at all times. We can speculate that in nature most of these quantities also fluctuate to some degree. The core accretion mechanism should therefore be able to produce planets of an even greater diversity than found here. On the other hand, the question of whether the core accretion can explain \textit{all} the planets (cf. Matsuo et al. \cite{matsuoetal2007}) is a much more difficult one. The models are probably not yet mature enough and observations are still too incomplete to allow a definitive statement.

Figure \ref{fig:aM} reveals that the synthetic planets are not randomly distributed inside the mass-distance plane. Instead, various concentrations, bars and depleted regions can be distinguished. One can also study the form of the quite well defined envelope filled by the synthetic population, and why certain regions remain empty. When comparing fig. \ref{fig:aM} with actual discoveries, one should however bear in mind the incompleteness of the model, affecting in particular low mass planets (\S  \ref{subsect:limitationsofthemodel}).

\subsubsection{The limiting envelope}\label{subsubsect:thelimitingenvelope}
At large masses, the planetary population is bounded by the largest  overall mass for a given semimajor axis, $\mmax (a)$. Outside $a\sim 3$ AU, $\mmax$ is a decreasing function of $a$, falling to about 100 $\mearth$ at $\sim20$ AU.  The reason is that the time needed to build up a massive core at large distances becomes comparable and ultimately longer than the gas disk dispersion timescale (Ida \& Lin \cite{idalin2004a}). See sect. \ref{subsubsect:type2variation} for processes which could modify this behavior.

Inside 3 AU, the behavior of $\mmax$ is inverted, \textit{i.e.} $\mmax$ is an increasing function of $a$. This means that there is an absence of very massive planets at small orbital radii. This is what has been discovered in the observed extrasolar planet population if only planets around single host stars are considered (Udry et al. \cite{udryetal2003}; Zucker \& Mazeh \cite{zuckermazeh2002}).  With the initial surface density profile used here ($\Sigma \propto a^{-3/2}$), we find that  inside 3 AU, $\mmax$ scales approximately as $a^{3/4}$ (as $\miso$), provided that type I migration is slow, see \S \ref{subsubsect:f1variation}. For populations obtained with higher type I migration rates  $\mmax$ is flat inside $\sim3$ AU. Future discoveries of a very large number of single giant planets out to several AU (Ge et al. \cite{geetal2007}) around single stars will help to define better the exact shape of $\mmax (a)$.

\subsubsection{Structures in the $a-M$ diagram}
The different phases of planet formation and migration that were identified  in the formation tracks leave traces also in the final $a-M$ of the planets. One can distinguish the ``failed cores'', the ``horizontal branch'', the  ``main clump'', and the ``outer group'' planets. As a new feature, fig. \ref{fig:aM} however also shows a depletion of planets with masses between 30 to 100 $\mearth$. This is the analogue of the  ``planetary desert'' first discussed by Ida \& Lin (\cite{idalin2004a}). Compared to their results, the depletion is much less severe in our simulations. 

The reason for this difference is difficult to pinpoint, as both formation models differ in many aspects, but it is at least partially due to the way the maximal gas accretion rate of the planets is calculated. Both models use the criterion that the gas accretion rate given by the planet's Kelvin-Helmholtz timescale (which we implicitly obtain by the structure calculations) must be smaller than the mass transfer rate in the disk. Ida \& Lin (\cite{idalin2004a}) however use this criterion only if the mass of the planet is additionally larger than the local gas isolation mass calculated from the unperturbed disk profile. This quantity is usually clearly larger (Ida \& Lin \cite{idalin2004a}) than the minimal planet mass required to fulfill the viscous and the thermal condition (Lin \& Papaloizou \cite{linpapaloizou1985}).  At these masses, the planet becomes able to tidally open a (partial) gap in the protoplanetary disk (e.g. Ida \& Lin \cite{idalin2008}). This could reduce the amount of gas which is effectively in the planet's direct gravitational reach, \textit{i.e.} its gas isolation mass.  Therefore, it seems possible that the planet mass where the accretion rate in the disk becomes the limiting factor might be significantly smaller than the gas isolation mass as assumed in Ida \& Lin (\cite{idalin2004a}), so that we disregard the second additional criterion, and always limit the planet's gas accretion rate to $\dot{M}_{\rm disk}$.  

As illustrated by the prototypical planet of the ``main clump'', the time at which planets start runaway gas accretion occurs for typical initial conditions generally at quite advanced stages of disk evolution. By then, the disk has undergone significant mass loss, and $\dot{M}_{\rm disk}$ is low. Therefore, disk limited planetary accretion rates are usually down to a few $10^{-4}\mearth$/yr so that growing a Jupiter-mass object requires an amount of time comparable to the remaining disk lifetime (see \S \ref{subsubsection:photoevaporation} for a population with very long, unrealistic disk lifetimes). Hence, the probability that the disk disappears while the planet is at an intermediate mass is not vanishingly small. This populates the ``planetary desert'' with intermediate mass objects, so that we predict only a moderate decrease of the relative number of planets with  masses between 30 to 100 $\mearth$ (about a factor 2-3 compared to Jovian planets, see the planetary initial mass function in paper II).  We have investigated this point by synthesizing a test population where we limit the gas accretion rate of a planet by $\dot{M}_{\rm disk}$ only if its envelope mass is larger than the gas isolation mass, as Ida \& Lin (\cite{idalin2004a}), and found that this results in a population that has indeed a stronger depletion of intermediate mass planets, which are then about 5 to 6 times less frequent than Jovian planets.

Long baseline, high precision RV surveys will test for the existence of the ``horizontal branch'' and the ``planetary desert'', and determine their extension and intensity. The upper mass limit of the branch constrains the minimal mass needed to go into runaway gas accretion, while  the lower mass boundary of the branch constrains the speed of type I migration.  The degree of a depletion of intermediate mass planets constrains planetary accretion rates during runaway gas accretion.

In the past 12 years, a significant number of extrasolar planets with masses much larger than one Jupiter mass were discovered. We also find such planets ($M\gtrsim10$ $\mj$) in our model, mainly in the ``outer group'', as shown by fig. \ref{fig:aM}. Their formation is the consequences of the fact that we do not reduce gas accretion due to gap formation. About 0.4 \% of all initial conditions lead to the formation of planets with a mass exceeding 12-13 $\mj$, which is commonly regarded as the brown dwarf limit (Chabrier et al. \cite{chabrieretal2000}). A handful of planets (out of the $\sim$50\,000) even reach a mass between 20 to 40 $\mj$. As it was shown recently that such objects burn deuterium in the layers above the core (Baraffe et al. \cite{baraffeetal2008}), they form an interesting population of deuterium burning planets. 

The synthetic population does not contain analogs of Uranus and Neptune both in term of mass and semimajor axis. Indeed, no planets at all are found outside $\sim20$ AU. Partially, this is simply an artifact of the lack of planet formation after the dispersion of the gas disk in our model(\S \ref{limitationslatetimeevo}). Uranus and Neptune are however insofar challenging that these planets accreted significant hydrogen-helium envelopes \textit{i.e.} that they were apparently formed while the gas disk was still present (Goldreich et al. \cite{goldreichetal2004a}; Chambers \cite{chambers2006}). 

It is well known that core accretion requires for the formation of Uranus and Neptune at their current position formation timescales which exceed typical gas disk lifetimes by a large factor (Pollack et al. \cite{pollacketal1996}; Thommes et al. \cite{thommesetal2003}). Hence, it was proposed that for Uranus and Neptune, one must give up the principle of a in-situ formation and take into account the possibility of an ejection because of N-body interactions with the giant planets (Thommes et al. \cite{thommesetal2002}; Tsiganis et al. \cite{tsiganisetal2005}). The fact that our population contains in the ``horizontal branch'' a significant number of planets with a mass and composition as the ice giants, but just at smaller orbital distances, goes well this second interpretation, and the general idea that planetary systems start with more crowded and compact configurations (e.g. Ford \& Chiang \cite{fordchiang2007}).

\subsection{Non-nominal populations}\label{subsect:nonnominalpopulations}
The representative tracks we have identified in the previous sections and the final $a-M$ were all calculated for one particular population, \textit{i.e.} a particular choice of underlying assumptions and parameters of the model like $\alpha$ or the type I migration efficiency factor. Here we discuss the effect of changing some settings. 

\begin{figure*}
\begin{minipage}[b]{.5\linewidth} 
\includegraphics[width=\textwidth]{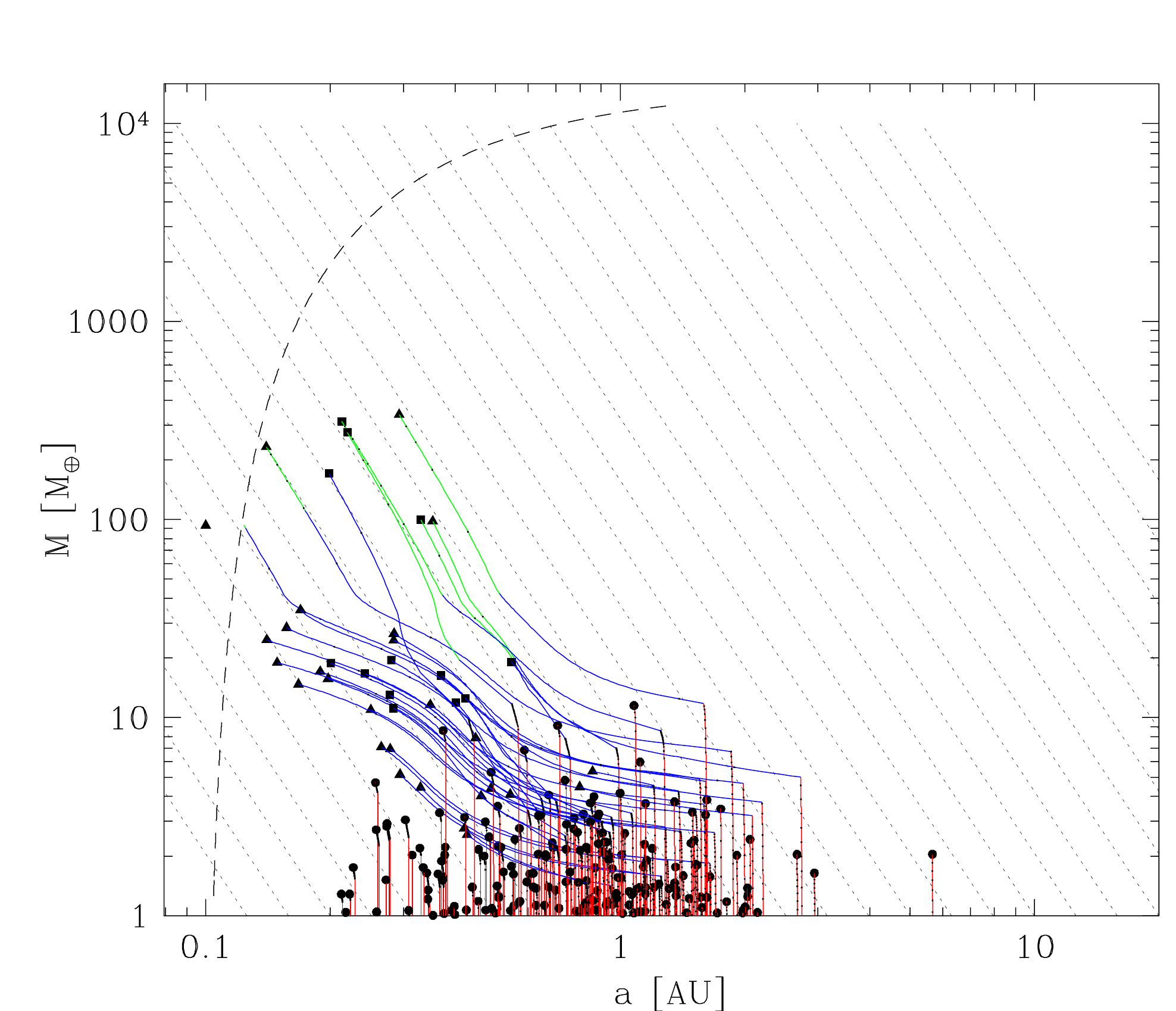}
\includegraphics[width=\textwidth]{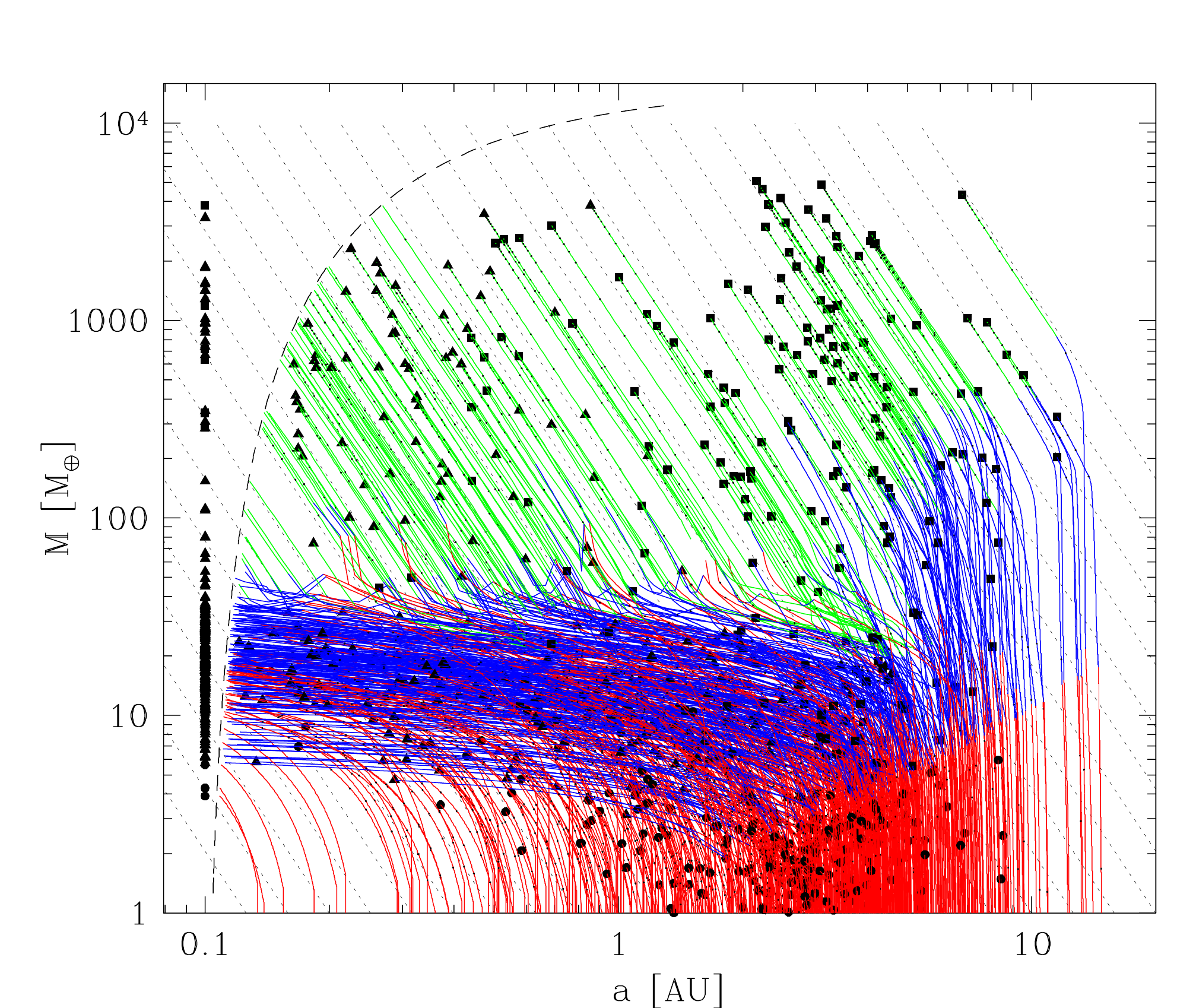}
\end{minipage}
\hspace{.0\linewidth}
\begin{minipage}[b]{.5\linewidth} 
\includegraphics[width=0.95\textwidth]{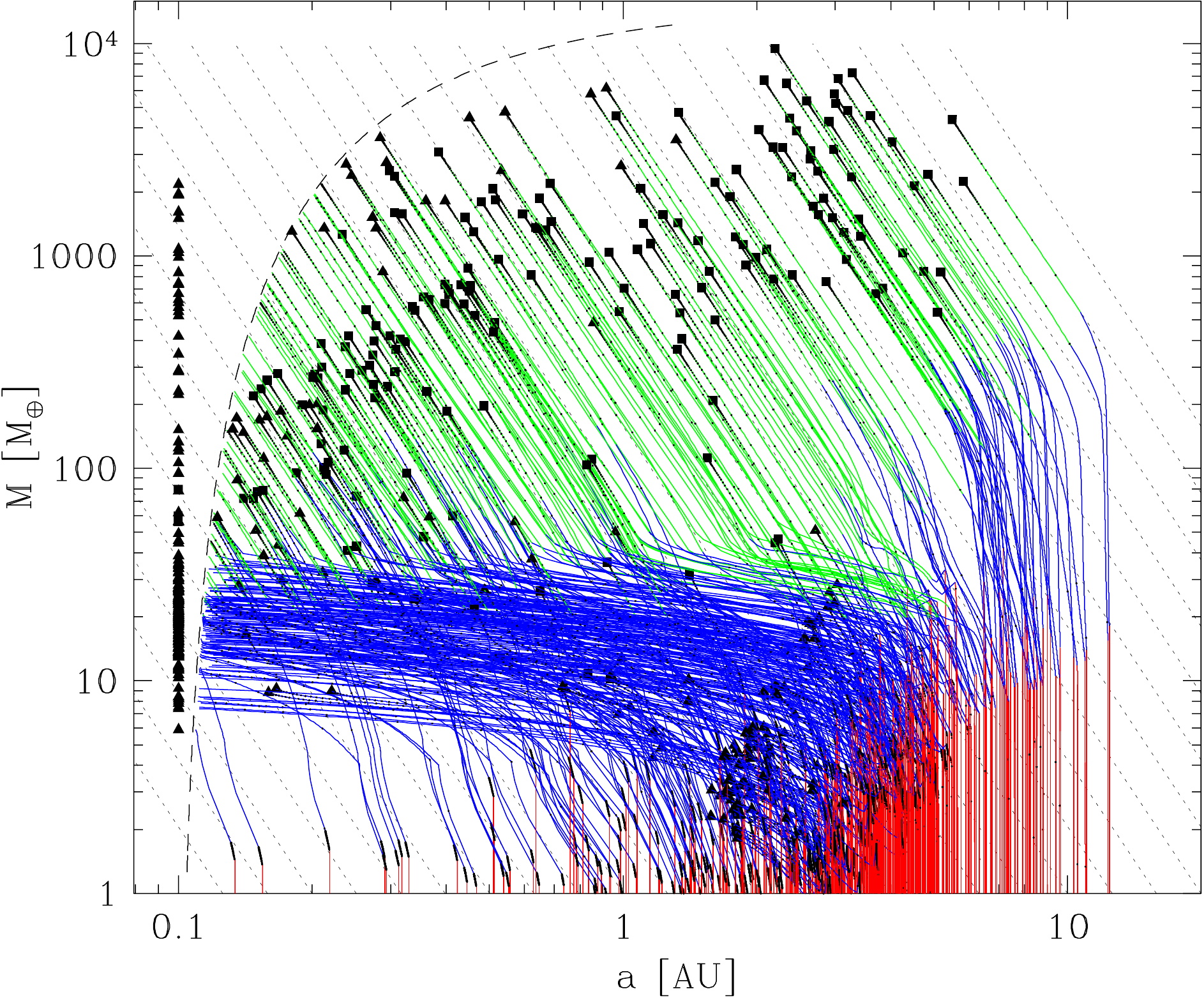}
\includegraphics[width=\textwidth]{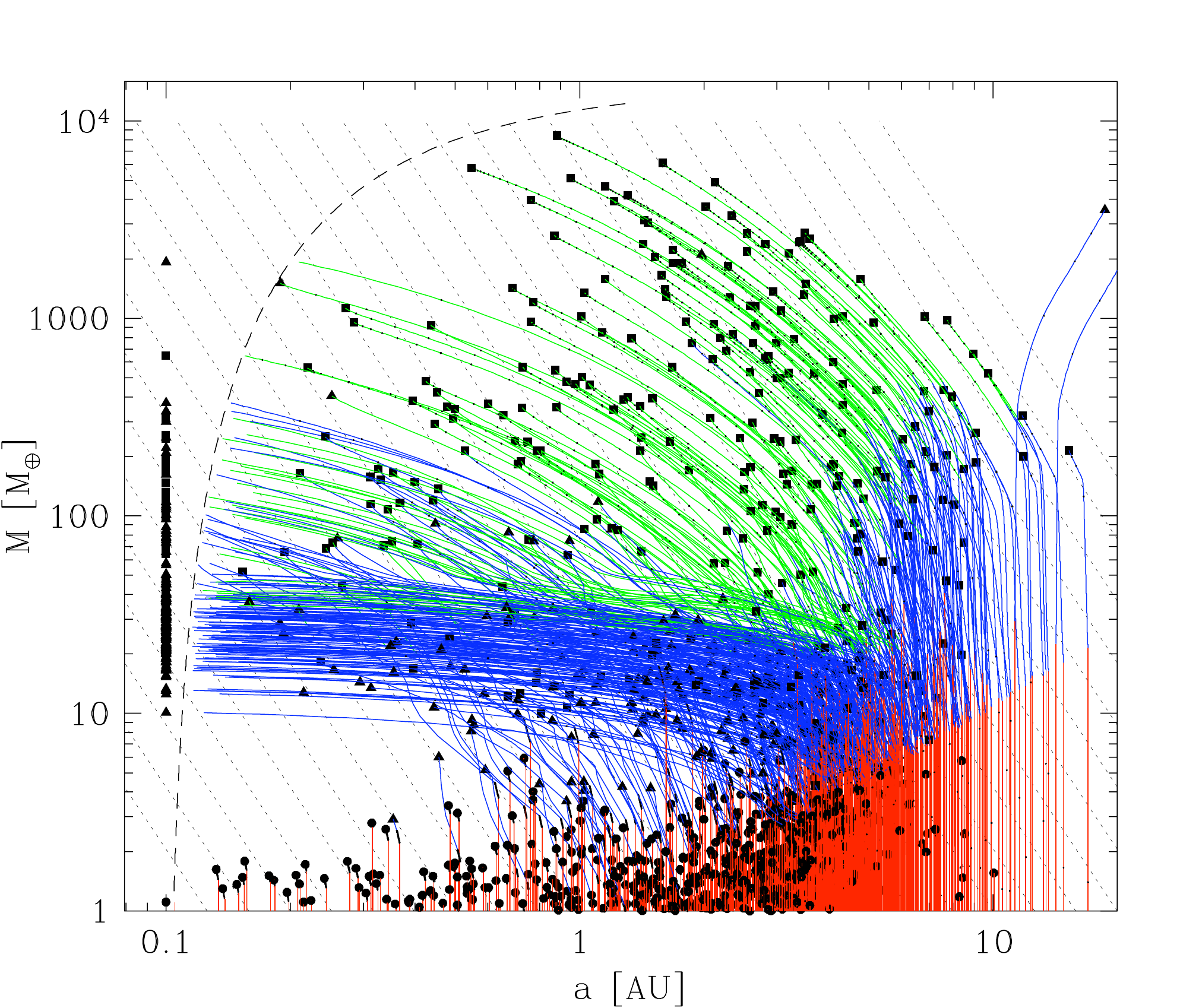}
\end{minipage}\caption{Planetary formation tracks as in fig. \ref{fig:aMtracks} but for non-nominal populations. Top left:  
For planetesimal eccentricities and inclinations as in Thommes et al. (\cite{thommesetal2003}). Top right: For $\mwind=0$ (no photoevaporation). Bottom left: For a type I migration efficiency factor $\f1$=0.1. Bottom right: For a type II migration rate as in Ida \& Lin (\cite{idalin2004a}). }
\label{fig:nonnominaltracks}
\end{figure*}

\subsubsection{Core growth regime}\label{subsubsection:coregrowthregime}
The accretion rate of planetesimals is crucial not only for the growth of the cores themselves, but indirectly also for the accretion of the envelopes. The solid accretion rate is a function of the velocity dispersion $v_{\rm disp}$ of the planetesimals. We use the prescription of Pollack et al. (\cite{pollacketal1996}) for $v_{\rm disp}$ corresponding to a situation between the shear and the dispersion dominated regime. Different results have been obtained concerning the importance of these two regimes (Ida \& Makino \cite{idamakino1993}; Rafikov \cite{rafikov2003}; Goldreich et al. \cite{goldreichetal2004b}). In the dispersion dominated picture of Thommes et al. (\cite{thommesetal2003}) planetesimal random velocities are higher than in the Pollack et al. (\cite{pollacketal1996}) description. Fortier et al. (\cite{fortieretal2007}) have studied the effects of  these high  $v_{\rm disp}$ on giant planet formation and found an increase of the formation time of Jupiter by around one order of magnitude relative to Pollack et al. (\cite{pollacketal1996}), depending on the mass of the disk.

We have synthesized a population where we follow Fortier et al. (\cite{fortieretal2007}) in calculating the planetesimal eccentricity  and inclination as in Thommes et al. (\cite{thommesetal2003}, their eq. 10).  The top left panel of fig. \ref{fig:nonnominaltracks} shows the resulting formation tracks and makes clear that there is a very strong effect, especially at large distances. Most seeds do not even grow from 0.6  to 1 $\mearth$. Only an extremely small fraction of initial conditions (extremely high $\fpg, \sigmanorm, \tdisk$) allows the formation of giant planets which are all inside 0.5 AU and have low masses. It is obvious that Kolmogorov-Smirnov tests (paper II) indicate a negligible probability that this synthetic population and the observed one come from the same parent distribution. We therefore conclude that in order to reproduce the observed extrasolar planets by the core accretion mechanism, solid accretion must occur on timescales clearly shorter than predicted by the model of  Thommes et al. (\cite{thommesetal2003}). Indeed, including additional effects like planetesimal fragmentation (Chambers \cite{chambers2006}; Kenyon \& Bromley \cite{kenyonbromley2009}) leads to significantly reduced core growth and planet formation timescales (see e.g. Thommes et al. \cite{thommesetal2008}). These effects will have, at least qualitatively, similar consequences as the low  $v_{\rm disp}$ we assume.

\subsubsection{Photoevaporation}\label{subsubsection:photoevaporation}
Another population was synthesized with photoevaporation switched off ($\mwind=0$), so that the disks evolve only due to viscosity. As expected, disk lifetimes are clearly increased for such a case. The mean disk lifetime is now about three times larger than in the nominal population. This is clearly incompatible with the observed distribution of disk lifetimes (\S \ref{subsection:photoevaporationrate}). The top right panel of fig. \ref{fig:nonnominaltracks} shows that the final positions of the planets in the $\mwind=0$ population differ significantly from the nominal case, even if the tracks leading there are similar. As expected, migration becomes more important, and a new group of massive planets inside about 0.5 AU is formed. The fraction of embryos that reach the inner boarder of the computational disk increases by about a factor three compared to the nominal case.  The fraction of  initial conditions that leads to giant planets increases too, as planets have more time to grow which also explains why a much emptier ``planetary desert'' is seen than in the nominal case. Statistical tests (paper II) show that this synthetic population is not compatible with the observed one: The KS tests for the $\msini$ and the semimajor axis distributions indicate a probability of only a few percent that the two samples come from the same parent population: The synthetic planets are too massive and too close-in. We therefore conclude that a distribution of disk lifetimes that is incompatible with the observational data (Haisch et al. \cite{haischetal2001}) leads to a synthetic population that is incompatible with the observed exoplanets, too. 

\subsubsection{Type I migration}\label{subsubsect:f1variation}
 Motivated by the short migration timescales found by Tanaka et al. (\cite{ Tanaka}) which indicate that cores are lost to the star before they can grow large enough to trigger runaway gas accretion, several authors have been re-examining type I migration rates and found considerable cause for uncertainties  (e.g. Nelson \& Papaloizou \cite{NelsonPap04};  Paardekooper \& Mellema \cite{paardekoopermellema2006}). Even if it is clear that the simple multiplication of  the migration rates of Tanaka et al. (\cite{Tanaka}) by a constant factor $\f1$  is only a first order approximation at best of the true type I migration rate (Kley \& Crida \cite{kleycrida2008}), the Monte Carlo simulations represent a suitable way to study the effects of various magnitudes of type I migration on the population as a whole, thereby possibly ruling out some values for $\f1$ (Ida \& Lin \cite{idalin2008}). 

We have therefore synthesized populations using also $\f1$=0.01, 0.1 and 1.0, besides the value of 0.001 used above. In fig. \ref{fig:nonnominaltracks}, bottom left, planetary formation track are plotted for a population using $\f1=0.1$ and otherwise nominal settings. For the high migration rates, one should keep in mind that migration is neglected before the seed embryo is put into the disk thereby the true migration is  partially underestimated (Ida \& Lin \cite{idalin2008}). 

At $\f1=0.1$, migration causes many ``failed cores'' to migrate into the feeding limit, especially at small distances as found also by Ida \& Lin (\cite{idalin2008}). A significant number of ``failed cores'' nevertheless remains at intermediate distances, due to the disappearance of the nebula. Varying the migration rate has also related, important consequence on the occurrence of close-in ($\lesssim 0.3$ AU) planets with masses $3\lesssim  M\lesssim10$ $\mearth$. Such planets do not exist for a low migration rate (cf. figs. \ref{fig:aMtracks} and \ref{fig:aM}) but become increasingly more numerous and smaller with increasing type I rate. This absence is the result of the combination of two effects. First, our model does not include the in-situ formation of terrestrial planets after disk dissipation. Hence, the formation of Mercury type planets and larger counterparts, or other mechanisms that lead to the formation of close-in terrestrial planets (see Raymond et al. \cite{raymondetal2007} for an overview) are not taken into account. This in-situ accretion would fill the empty area ``from below'' (in mass). Due to the limited amount of solids very close to the star, the mass to which planets can grow in-situ is however limited. Second, in most disks seed embryos can only start  beyond the iceline (cf. the distribution of $\astart$, \S \ref{subsect:embryostarposition}). Thereafter, they must be brought in by migration without growing too much in order to have a small final mass. This is only possible for fast migration (compared to accretion), and sets a minimal mass of planets that fill the empty region ``from outside'' (in semimajor axis).

For $\f1=0.1$ and 1.0, the mass of planets that migrate to $\atouch$ is usually larger than about 6 and 2 $\mearth$, respectively. For all $\f1$, a few planets with even smaller masses are also found inside the feeding limit which come from very small initial positions $\astart$. It should further be noted that this result depends on the assumed radial surface density profile of solids. The  $\sigmad\propto a^{-3/2}$ we use for simplicity down to 0.1 AU leads to high solid surface densities near the star. For a truncated disk profile used in tests, which might be more appropriate, and $\f1=0.1$, even planets as small as $\sim2$ $\mearth$ have reached the feeding limit, and the region inside $\sim0.3$ AU which is empty at low $\f1$ is then filled.

Observations able to detect planets in the range 1 to 10 Earth masses close to the star will show if there is a minimal mass for planets inside a few tens of an AU, and whether the mass spectrum there is continuous, or contains a gap between planets formed in-situ and cores brought in from further out by migration. Hence, such object will potentially provide us with an indicator on the efficiency of type I migration. 

Considering the very massive planets  ($\gtrsim 10$ $ \mj$), it is found that at low type I migration rates, no such planets are found within about 0.4 AU of the star. At high rates, some very massive planets have, in contrast, migrated to the feeding limit.  It is interesting to note that the type I migration, which affects directly small planets, also has a significant and measurable effect on the most massive planets as it allows planets forming at small distances to have a quicker access to larger amounts of solids than the locally available $\miso$ for $\f1=0.001$.

\subsubsection{Type II migration}\label{subsubsect:type2variation}
The formation tracks of giant planets of both of the ``main clump'' and the ``outer group'' show the importance of the planet-dominated migration phase, as well as the disk limited gas accretion rate (which itself results from the process of gas transport inside the protoplanetary disk) in the braking phase. It is interesting to note that, using planet formation models following the growth and migration of a large number of embryos, Thommes et al. (\cite{thommesetal2008}) came to the same conclusion regarding the importance of these two processes.

The braking phase plays a crucial role for the final $a-M$ position of gas rich planets. It is clear that our description of the planetary $dm/dt$ and $da/dt$ in this phase remains a first approximation that could be significantly improved by a more sophisticated description of planet-disk interactions. For example, the eccentric instability that justifies the assumption that the planetary gas accretion rate is the same as the disk accretion rate occurs only for planets larger than a certain mass (Kley \& Dirksen \cite{kleydirksen2006}). At smaller masses, the accretion rate could be smaller (D'Angelo et al. \cite{dangeloetal2002}). Also planet-planet interactions of several giant planets forming concurrently can have important consequences on the migration and accretion rate. This was shown by Thommes et al. (\cite{thommesetal2008}), where a chain of migrating giant planets, locked together by mean motion resonances, is a frequent configuration. Then, only the outermost planet directly exchanges torques with the gas disk, and accretes gas from it. While with our one embryo per disk approach, we obviously cannot model such a behavior, we note that this configuration typically leads to an even earlier transition to the planet-dominated regime as multiple planets are better at holding off the outer disk than one (Thommes et al. \cite{thommesetal2008}).       

The  $da/dt$ of  a satellite that is massive compared to the local disk mass was studied by Syer \& Clarke (\cite{syerclarke1995}) and Ivanov et al. (\cite{ivanovetal1999}).  They both agree that in this situation, the migration timescale becomes longer than the viscous timescale,  but quantitatively their results differ.
The differences are linked to the question of what part of the angular momentum flux is effective in  moving the planet (Ida \& Lin \cite{idalin2005}), and whether gas overflows the gap formed by the planet (Armitage \cite{armitage2007}). Therefore, there is some uncertainty on the exact migration rate in the case of a massive planet. 

To have an estimate of the sensitivity of our results on the migration rate in the braking phase we have synthesized a test population replacing eq. \ref{eq:migrationrateduringbraking} by the prescription of Ida \& Lin (\cite{idalin2004a}). The bottom right panel of  fig. \ref{fig:nonnominaltracks} shows the resulting formation tracks. One sees that there is a visible modification of the tracks, but that the basic types remain. The comparison of the tracks with the dashed lines (slope equal $-\pi$) shows that with this alternative prescription, type II migration occurs on a shorter timescale. The reason is that the $da/dt$ is now proportional to $\Sigma(R_{m})R_{m}^2$ instead of $\Sigma(\aplanet)\aplanet^2$, where $R_{m}$ is the radius of maximum viscous couple, which moves outwards very quickly, and is in almost all cases larger than $\aplanet$ once planets are large enough to migrate in type II. Therefore is the first quantity usually larger than the second one (fig. \ref{fig:sigmaaalocaldiskmass}). 

There are also two examples of tracks where outward migration occurs. As expected from the quick outward movement of  $R_{m}$, only a very small group of about 10 seeds (out of $\sim10000$ initial conditions) has undergone significant outward migration\footnote{Note that we calculate in this test population $R_{m}$ as Ida \& Lin (\cite{idalin2004a}), and not in a self-consistent way from the disk model, where photoevaporation can influence the evolution of $R_{m}$ at late times and hence the outward migration (Veras \& Armitage \cite{VA04}).}. The maximal planetary mass $\mmax$ at $\sim20$ AU is therefore increased relative to the nominal model (sect. \ref{subsubsect:thelimitingenvelope}) by about one order of magnitude.

The discovery of many more giant planets outside $\sim20$ AU by techniques like astrometry or direct imaging (Kalas et al. \cite{kalasetal2008};  Marois et al. \cite{maroisetal2008}) would indicate that besides outward migration  one or several of the following mechanisms not included in the nominal model is important:  Scattering of planetary seeds to large $\astart$ early during formation (``monarchical growth'',  Weidenschilling  \cite{weidenschilling2005}, \cite{weidenschilling2008}), a completely different formation mechanism (direct gravitational collapse, e.g. Boss \cite{boss2001}; Mayer et al. \cite{mayeretal2004}) or scattering of planets in initially crowded multiple planetary systems after formation (e.g. Veras \& Armitage \cite{VA04}). Note however for the last possibility that the simulations of Thommes et al. (\cite{thommesetal2008}), which combine in a self-consistent way the formation and subsequent N-body interactions, do not usually lead to the formation of giant planets further out than $\sim20$ AU.  In another non-nominal population, we have addressed the early embryo ejection and used a modified prescription to calculate the starting time $\tstart$ that approximately mimics  a ``monarchical  growth'' mode (Weidenschilling \cite{weidenschilling2005}, \cite{weidenschilling2008}). In this case, while leaving the population in the inner system relatively untouched, a population of very massive planets outside $3$ AU comes into existence so that the maximal planetary mass $\mmax$ does not decrease anymore outside $\sim3$ AU, but remains constant out to $\sim 10-20$ AU. 

\subsubsection{Transition from type I to type II migration}\label{subsubsect:transitiontypeItypeII}
As mentioned at the beginning of the paper, the transition from type I to type II occurs when the planet can open a gap in the protoplanetary disk. In our nominal model we assume that this occurs when the disk scale height becomes smaller than the planet's Hill radius (which is the relevant criterion in the limit of very large Reynolds numbers). In order to infer the influence of this part of our model, we have synthesized non-nominal populations using the type I/type II transition condition derived by Crida et al. (\cite{cridaetal2006}). Using the low type I migration of the nominal model, only very few ``Hot'' planets are then formed, since the transition to (faster) type II occurs at higher masses. However, when $\f1$ is increased, we obtain similar formation tracks and final  sub-populations. Note that in this case, very few planet migrate in the disk-dominated type II regime, and a majority of them passes directly from type I to the planet-dominated type II regime. We mention finally that, as outlined by e.g. Armitage and Rice (\cite{armitagerice2005}), the transition, and especially the migration rate during the transition, is not fully  understood and could result in special migration regimes not considered here (e.g. Papaloizou \& Terquem \cite{papaloizouterquem2006}). This incomplete understanding of migration in general remains one of the major uncertainties in planet formation theories of today.

\section{Summary and conclusion}\label{sect:conclusions}
We have presented our extrasolar planet population synthesis calculations.  As formation model we use a slightly simplified version of the extended core accretion model presented in Alibert et al. (\cite{alibertetal2005a}) which has been shown to reproduce many observed properties of the giant planets of our own solar system (Alibert et al. \cite{alibertetal2005b}). 

We use four random variables to describe the possible initial conditions for planet formation: The dust-to-gas ratio $\fpg$, the initial gas surface density $\sigmanorm$, the photoevaporation rate $\mwind$ and the starting position of the embryo in the disk, $\astart$. The distributions  for the first three Monte Carlo variables can be derived from observed properties of stars of the solar neighborhood, or protoplanetary disks.  For the dust-to-gas ratio, we use the distribution of [Fe/H] of the stars in the CORALIE planet search sample (Santos et. \cite{santosetal2003}). For the gas surface densities, we use the disk mass distribution in $\rho$ Ophiuchi (Beckwith \& Sargent \cite {beckwith1996}). For the photoevaporation rate, we use the distribution of disk lifetimes by Haisch et al. (\cite{haischetal2001}). We have also discussed the complications that arise from the conversion of such observed quantities into figures that can be used in numerical simulations. The last random variable, $\astart$ cannot be derived from observations. Here we follow Ida \& Lin (\cite{idalin2004a}) and use a distribution that is uniform in $\log(a)$.

With the adopted random distributions, we find that our formation model predicts a population of synthetic planets that is very diverse, not unlike the actually observed population. The final semimajor axis varies by two orders of magnitude, and the mass by four orders. Similarly, the internal bulk composition is very different, and covers gas giants but also ``Super Earth'' mass planets with a envelope to core mass ratio similar to Venus. We conclude that this diversity illustrates the ability of the underlying concepts of core formation to serve as an unified formation model for planets of masses ranging from a few times the Earth mass to beyond the brown dwarf limit. The observed diversity of extrasolar planet is a natural consequence of the different properties of protoplanetary disks.

Planet formation is a process of concurrent mass accretion and migration which can be well represented by planetary formation tracks in the mass-distance plane. These tracks show a number of distinct phases, which are visible in the final $a-M$ distribution of the planets, too.

In the first phase, at small masses, planets migrate in type I migration (which must however be slow to be compatible with observations,  cf. paper II),  and accrete mostly solids. Planets which remain in this stage until the moment when the disk disappears form a vast sub-population of low mass, core-dominated planets, ``failed cores''. The distributions of protoplanetary disks properties are such that this is the most likely outcome of the formation process. This is consistent with the observation that 90 to 95\% of FGK stars in the solar neighborhood apparently remained without giant planets. 

If the disk properties are instead such that the planet grows massive enough to open a gap in the disk, the second phase starts. Then, planets migrate inwards in type II migration, collecting solids on their way in. In this phase, migration occurs on a shorter timescale than accretion, so that the tracks of the planets show a ``horizontal branch'' phase in the $a-M$ plot. Planets in this phase have masses between 10 to 30 $\mearth$, a static gaseous envelope and an internal gas to solid ratio similar to Neptune. 

Some planets on the ``horizontal branch'' collect  enough solids to become supercritical for gas runaway accretion. Their gas accretion rate then quickly reaches the disk limited value, and their mass becomes large compared to the local disk mass. Therefore, giant gaseous planets reach their final mass and semimajor axis in the third phase, when they accrete gas at the rate controlled by the disk  and migrate in the planet dominated, slower type II regime.  This leads to the formation of a concentration of giant planets between 0.3 to 2 AU in the ``main clump'', with masses between $\sim100$ $\mearth$ and 3 $\mj$. 

Embryos starting far from the parent star in metal rich disk can grow supercritical for gas runaway accretion in-situ, without passing through the ``horizontal branch''. This leads to the formation of an ``outer group'' of giant planets. 

Apart from these four sub-populations, we find in the final nominal $a-M$ diagram (1) an absence of massive  ($\gtrsim 10\mj$) planets both close ($\lesssim 0.5$ AU) to the star and very far ($\gtrsim 10$ AU) from it, (2) a certain depletion of planets with masses between 30 and 100 $\mearth$, in analogy to the ``planetary desert'' (Ida \& Lin \cite{idalin2004a}) which is however not very strong (a factor 2-3 relative to giant planets), as at the time planets start runaway gas accretion, the disk limited accretion rate, which is decisive in this regime, is usually already quite low, (3) a  handful very massive ($\gtrsim 20 \mj$) deuterium burning planets (Baraffe et al. \cite{baraffeetal2008}), showing that this mass domain can be populated by different formation channels, (4) an absence of Neptune analogs in terms of \textit{both} mass and semimajor axis, but many such planets at smaller distances.

Population synthesis is a valuable tool to asses the global consequences of model settings. From the synthesis of a number of non-nominal populations we find that (1) cores must form quickly (Thommes et al. \cite{thommesetal2003}; Chamber \cite{chambers2006}), (2) disks with lifetimes too long compared to observations lead to too massive planets too close-in, (3) fast type I migration brings low mass planets close to the parent star, so that the depletion of planets with $3\lesssim M/\mearth \lesssim 10$  inside 0.3 AU in the nominal population vanishes, (4) the early ejection of seeds, and outward type II migration could result in very massive planets out to about 20 AU, but only in small numbers,  (5) the transition regime between type I and type II migration has important implications for the final population.

With improving detection methods leading to a rapidly increasing number of known extrasolar planets, the actual distributions of masses and semi-major axis can be determined with growing statistical confidence and compared to model predictions. Hence, planet populations studies are the ideal tools to extract from these distributions constraints on various aspects of the formation models thereby reaping the scientific benefits of the large observational efforts invested. In the companion paper (paper II), we carry out a direct statistical comparison between a carefully chosen sample of actual exoplanets and detectable synthetic planets and use this comparison to extract relevant constraints on the formation mechanism. Last but not least, population synthesis can itself be used to guide and perhaps optimize the next generation of detection instruments.

\acknowledgements
We thank Stephane Udry and Shigeru Ida for useful discussions. We are thankful for detailed and interesting comments by two anonymous referees. This work was supported in part by the Swiss National Science Foundation. Computations were made on the ISIS, ISIS2 and UBELIX clusters at the University of Bern, and on the cluster of the Observatoire de Besan\c{c}on funded by the Conseil G\'en\'eral de Franche-Comt\'e.

\end{document}